\documentclass[aip,jcp]{revtex4-1}
\usepackage{dcolumn}
\usepackage{bm}

\usepackage[utf8]{inputenc}
\usepackage[T1]{fontenc}
\usepackage{mathptmx}
\usepackage{etoolbox}
\draft 
\usepackage{color}
\usepackage{graphicx}
\usepackage{xr}

\makeatletter
\def\@email#1#2{%
 \endgroup
 \patchcmd{\titleblock@produce}
  {\frontmatter@RRAPformat}
  {\frontmatter@RRAPformat{\produce@RRAP{*#1\href{mailto:#2}{#2}}}\frontmatter@RRAPformat}
  {}{}
}%
\makeatother

\makeatletter
\newcommand*{\addFileDependency}[1]{
  \typeout{(#1)}
  \@addtofilelist{#1}
  \IfFileExists{#1}{}{\typeout{No file #1.}}
}
\makeatother

\newcommand*{\myexternaldocument}[1]{
    \externaldocument{#1}
    \addFileDependency{#1.tex}
    \addFileDependency{#1.aux}
}

\myexternaldocument{SI}
\begin{document}

\preprint{}

\title{Innate Dynamics and Identity Crisis of a Metal Surface Unveiled by Machine Learning of Atomic Environments} 


\author{Matteo Cioni}
\affiliation{ 
Department of Applied Science and Technology, Politecnico di Torino, \\Corso Duca degli Abruzzi 24, 10129 Torino, Italy
}%
\author{Daniela Polino}%
\affiliation{ 
Department of Innovative Technologies, University of Applied Sciences and Arts of Southern Switzerland, \\Polo Universitario Lugano, Campus Est, Via la Santa 1, 6962 Lugano-Viganello, Switzerland
}%
\author{Daniele Rapetti}
\affiliation{ 
Department of Applied Science and Technology, Politecnico di Torino, \\Corso Duca degli Abruzzi 24, 10129 Torino, Italy
}%

\author{Luca Pesce}
\affiliation{ 
Department of Innovative Technologies, University of Applied Sciences and Arts of Southern Switzerland, \\Polo Universitario Lugano, Campus Est, Via la Santa 1, 6962 Lugano-Viganello, Switzerland
}%
\author{Massimo Delle Piane}
\affiliation{ 
Department of Applied Science and Technology, Politecnico di Torino, \\Corso Duca degli Abruzzi 24, 10129 Torino, Italy
}%

\author{Giovanni M. Pavan}
\email{giovanni.pavan@polito.it}
\affiliation{ 
Department of Applied Science and Technology, Politecnico di Torino, \\Corso Duca degli Abruzzi 24, 10129 Torino, Italy
}%
\affiliation{ 
Department of Innovative Technologies, University of Applied Sciences and Arts of Southern Switzerland, \\Polo Universitario Lugano, Campus Est, Via la Santa 1, 6962 Lugano-Viganello, Switzerland
}%




\date{\today}

\begin{abstract}
Metals are traditionally considered hard matter. However, it is well known that their atomic lattices may become dynamic and undergo reconfigurations even well-below the melting temperature. The innate atomic dynamics of metals is directly related to their bulk and surface properties. Understanding their complex structural dynamics is thus important for many applications but is not easy. Here we report deep-potential molecular dynamics simulations allowing to resolve at atomic-resolution the complex dynamics of various types of copper (Cu) surfaces, used as an example, near the Hüttig ($\sim1/3$ of melting) temperature. The development of a deep neural network potential trained on DFT calculations provides a dynamically-accurate force field that we use to simulate large atomistic models of different Cu surface types. A combination of high-dimensional structural descriptors and unsupervised machine learning allows identifying and tracking all the atomic environments (AEs) emerging in the surfaces at finite temperatures. We can directly observe how AEs that are non-native in a specific (ideal) surface, but that are instead typical of other surface types, continuously emerge/disappear in that surface in relevant regimes in dynamic equilibrium with the native ones. Our analyses allow estimating the lifetime of all the AEs populating these Cu surfaces and to reconstruct their dynamic interconversions networks. This reveals the elusive identity of these metal surfaces, which preserve their identity only in part and in part transform into something else in relevant conditions. This also proposes a concept of “statistical identity" for metal surfaces, which is key for understanding their behaviors and properties.
\end{abstract}


\maketitle 

\section{Introduction}

Metals are traditionally considered a reference materials class in hard matter. Their structure, characterized by crystalline solid lattices with well-defined order and symmetries,\cite{desjonqueres1996concepts} imparts them a variety of technologically-relevant properties. Nonetheless, the dynamics of metals is perhaps as much interesting, but also far more elusive. 

It is known that metals may assume a non-trivial dynamic behavior well below their melting temperature. Above the H\"uttig temperature ($\sim1/3$ of the melting temperature), their atomic surface enters a dynamic equilibrium in which the atoms may move in the lattice.\cite{spencer1986stable,jayanthi1985surface,bernasconi1993reconstruction}

In particular it has been experimentally observed that above this temperature metal surfaces may undergo different structural transformations, such as "deconstruction", "preroughening", faceting, roughening and surface premelting. However, although a large number of experimental and theoretical studies focused on these phenomena in the past\cite{trayanov1990anisotropic,durr1991anomalous,merikoski1994disordering,zeybek2006thermal}, a general understanding is still far from complete.
Nonetheless, the appearance or disappearance of specific atomic environments and their dynamic evolution on a metal surface may be crucial in determining properties that are directly related to the structural features of their surface, such as, \textit{e.g.}, their reactivity in heterogeneous catalysis.\cite{xie2013catalysis,dattila2020active}

Obtaining atomic-level insight into the dynamics of metals is key to understanding their properties but is non-trivial.
Widely used for studying metals,\cite{yamakov2004deformation,zepeda2017probing,Wang2021,Norskov2009,Norskov2011,CalleVallejo2015,Zhong2020,Mavrikakis2021} computational modeling holds a considerable potential in this sense.
However, in several computational approaches, such as in heterogeneous catalysis studies, metal surfaces are usually treated as static.\cite{Norskov2009,Norskov2011,CalleVallejo2015,Zhong2020,Mavrikakis2021} Accounting for the atomic dynamics of metal lattices, nevertheless, becomes crucial in those conditions where this is determinant for the material's properties. 
As a notable example, Gazzarrini \textit{et al.} demonstrated how atoms mobility in copper (Cu) nanoparticles may produce variations in the number of vertex, edge and face atoms, affecting the nanoparticles efficiency in catalyzing CO$_2$ conversion to methane.\cite{gazzarrini2021born} 
Nelli \textit{et al.} studied the dynamic diffusion of atomic impurities in copper-cobalt icosahedral nanoparticles \textit{via} metadynamics simulations.\cite{Nelli2021}
These works offer important preliminary evidence of how metal lattices cannot be simply studied as static materials, but in certain regimes these are rather complex systems where atoms are in dynamic equilibrium and continuous exchange. 
Intriguingly, this is close to what it has been recently seen in soft self-assembling systems.\cite{crippa2022molecular,gasparotto2019identifying,gardin2022classifying,bochicchio2017into}  

While disentangling such complex structural dynamics is non-trivial, machine learning (ML) is useful to this end. 
On one hand, ML potentials trained on QM data offer the opportunity to obtain accurate force fields allowing to reliably simulate metals on relevant spatiotemporal scales. Since the pioneering work of Behler and Parrinello\cite{BehlerParrinello2007} that introduced the concept of high-dimensional neural network potentials, different approaches to build ML potentials have been developed in the last decade\cite{Csanyi2010,DeFabritiis2021TorchMD,SchNet2018,PhysNet2019,DeVita2015,deepMD,Pun2019-PINN,SNAP2020}, allowing to investigate with \textit{ab initio} accuracy systems of increasing complexity (we refer the reader to Ref. [\onlinecite{Unke2021ChemRev}] for a description of various widely used methods).
On the other hand, high-dimensional descriptors and advanced statistical analyses are extremely useful to unravel the complexity of dynamic molecular systems. For example, unsupervised clustering of smooth overlap of atomic positions (SOAP) data~\cite{SOAPBartok,SOAPDistance,capelli2022ephemeral} extracted from molecular dynamics (MD) simulations recently allowed to reconstruct the structural/dynamical complexity of a variety of molecular materials/systems\cite{gasparotto2019identifying,bian2021electrostatic,gardin2022classifying}
and to build robust data-driven metrics\cite{capelli2021data,capelli2022ephemeral} for their classification.\cite{gardin2022classifying}
Similar data-driven approaches have been also recently used to explore the mechanism of gold NPs melting.\cite{zeni2021data}
In this work, we report a data-driven approach that allows resolving at atomistic-resolution the complex dynamics of metal surfaces in relevant conditions. As an example of dynamic metal widely used for technological applications, we focus on Cu (the approach is nonetheless versatile and applicable to a variety of metals). 
An extensive campaign of DFT\cite{kohn1965self} MD simulations generated Cu surface configurations and interaction energy data that we used to train a deep neural-network (\textit{NN})-potential.\cite{deepMD} This provided a dynamically-accurate atomistic force field allowing to simulate various types of Cu surfaces near the H\"uttig temperature on relevant spatiotemporal scales. \textit{Bottom-up} and \textit{top-down} SOAP-based analyses reveal the atomic environments that statistically populate Cu surfaces, including non-native atomic environments, typical of other types of surfaces, which continuously emerge/resorb in the various Cu surfaces in equilibrium conditions.
This ultimately allows estimating an ``equivalent identity'' for such metal surfaces on a purely statistical basis from the data emerging along the DPMD simulations, changing the way we look at such materials.

\section{Results}
\subsection{ Modeling the dynamics of copper surfaces}

Here we use Cu as an example of dynamic metal widely used for various applications.\cite{nitopi2019progress,manthiram2014enhanced,hickman2012high, Magdassi_electronics,Lu2004Electric} 
We focus on the study of Cu surface models at 500-700~K, just above the H\"uttig temperature (447~K for Cu).
Noteworthy, although this is not central in this work, such conditions are of interest, \textit{e.g.}, for catalytic applications (CO$_2$ hydrogenation),\cite{behrens2012active,higham2020mechanism} where the structural/dynamical features of the Cu surface are important.

\begin{figure*}[ht!]
\centering
\includegraphics[width=\textwidth]{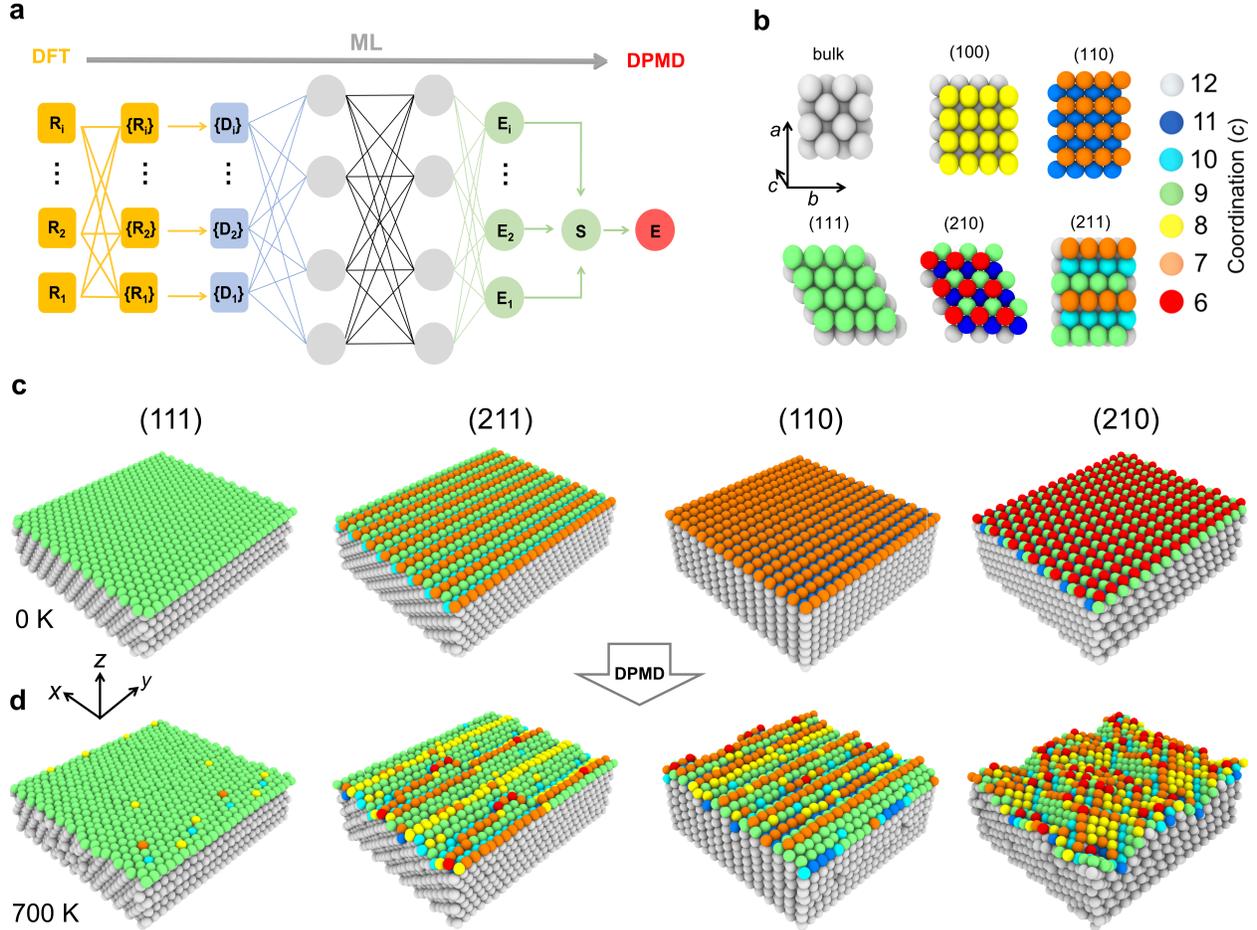}
\caption{\textbf{Atomistic DPMD simulations of Cu surfaces near the H\"uttig temperature.} \textbf{a} Scheme of the DeepMD\cite{deepMD} model developed in this work: Atomic configurations and energies obtained from DFT calculations have been used to train an inter-atomic \textit{NN}-potential for the Cu surfaces.
\textbf{b} Atomistic models for bulk and surface Cu environments (top view) used for the training (left: atoms are colored based on their coordination). DFT configurations and energies extracted in the range of temperature of 500-700~K have been used to train the Cu \textit{NN}-potential.
\textbf{c} Starting ideal configurations of the Cu surface models investigated in this work: we compare the behaviors of (111), (211), (110), and (210) Cu surfaces (atoms colored based on their coordination).
\textbf{d} Cu surfaces after 150 ns of DPMD simulations at 700~K show structural rearrangements, atomic mobility and coordination changes (left-to-right, the surfaces are ordered according to their mobility).
}
\label{fig1}

\end{figure*}

Studying the structural dynamics of metal surfaces requires sufficient accuracy in the treatment of the Cu-Cu interactions and, at the same time, one needs to simulate these atomic systems on sufficiently large spatiotemporal scales to prevent finite-size effects and guarantee that microscopic dynamic transitions (and not only vibrations) are sampled with sufficient statistics.
Accurate DFT\cite{kohn1965self} calculations are limited to timescales and sizes that are too short/small to this end, whereas classical atomistic force fields\cite{Gupta_energy,EAM1986,Rosato1989,MEAM1992,MEAM2003, MendelevEAM2008} may not guarantee sufficient accuracy in the treatment of the structural dynamics of the metal surface.
To obtain a dynamically-accurate atomistic force field, we turned to ML.
In particular, we developed an inter-atomic interaction potential for Cu surfaces \textit{via} training a deep \textit{NN} on data extracted from DFT calculations adopting the Deep Potential Molecular Dynamics (DeePMD) scheme developed by Zhang et al.\cite{deepMDPRL,deepMD} 
(Figure~\ref{fig1}a).

A campaign of DFT MD simulations of small Cu FCC bulk, (100), (110), (111), (211), and (210) surface environments conducted at temperatures between 500 and 700~K (Figure~\ref{fig1}b) provided a rich dataset of atomic configurations, forces and energies used to train a first guess \textit{NN}-potential. 
An active (iterative) learning strategy\cite{Deringer2017, dpgen} was then adopted to ensure a good sampling of the reconstructed configurations in the Cu surface and the local atomic transitions leading to such reconstructions. It is important to note that, although the surface portions simulated at DFT level and that are used for the training (Figure~\ref{fig1}b: $\sim100$ atoms, depending on the surface type -- sufficient to account up to the $\sim4-5$ atom neighbors) are smaller than the surfaces that are then simulated at atomistic level (Figures~\ref{fig1}c,d: containing $\sim2400$ atoms), the iterative nature of the approach and our tests demonstrate that the obtained potential is robust and reliable (see also Methods section for details). 
In fact, even if the first trained \textit{NN}-potential initially contains DFT-level information only on the small ideal surface patches (limited sampling), the new conformations that are then discovered \textit{via} deep-potential molecular dynamics (DPMD) simulations using such incomplete \textit{NN}-potential are then re-simulated at DFT-level and added to the DFT dataset. 
Such a process is conducted iteratively, and at each iteration the DFT data on the newly discovered atomic configurations are added to the training set. The training process ends when no new configurations are discovered in the successive iterations, and the obtained \textit{NN}-force field can be thus considered complete (the reduced spatial sampling is compensated by temporal sampling guaranteed by the iterative DPMD simulation scheme).

To ensure that no residual spurious finite-size effects could affect the trained \textit{NN}-force field, we also conducted additional tests  using larger surface patches than those of Figure~\ref{fig1}b ($\sim600$ atoms, six-times larger). Nonetheless, these tests demonstrated that the maximum deviation in the forces provided by the obtained \textit{NN}-potential in the two cases is negligible (estimated forces within the training error in the 99.95\%
of the cases), confirming the robustness and consistency of the obtained Cu \textit{NN}-potential.
We also note that the discovery-and-sampling approach adopted herein to create the training data-set is based on the collection and sampling of configurations along the DPMD simulations with a time-frequency suitable to effectively follow with fine-temporal resolution the motions of the atoms on the surfaces. In particular, the DFT training set contains information not only on the local minima configurations but also on the intermediate configurations and on the transition barrier states that are visited. In this way, the trained \textit{NN}-potential has DFT accuracy in reproducing both the energy differences and the transition barriers between the various atomic configurations visited along the DPMD simulations.
This provides a structurally and dynamically accurate force field having DFT precision in the treatment of the atomic configurations (energies, forces, etc.) and of their dynamic interconversion within the Cu surfaces (see Methods section for complete details).

The \textit{NN}-potential was finally validated against some copper bulk and surface properties. In particular we calculated the lattice constant, the vacancy and interstitial formation energies the surface energies at 0 K of the unreconstructed (100), (110), (111), (211) and (210). The values computed, reported in Table S1,   are in good agreement with the values computed at the DFT level, with the embedded atom model (EAM) of Mendelev et al.\cite{MendelevEAM2008}, and with available experimental values.\cite{kittel2018introduction,brandes2013smithells,TYSON1977267}
To ascertain the reliability of the \textit{NN}-potential, we calculated the surface energy of the (110) (1$\times$2) missing-row reconstruction at 0 K and found it to be larger than that of the unreconstructed (110) surface. In agreement with the experimental evidence that the Cu (110) surface at low temperatures does not undergo the (1$\times$2) missing-row reconstruction typical of other noble metals, \textit{e.g.} Au and Ag.  Lastly, we computed the diffusion barriers of single adatoms on the (100), (110) and (111). The results, reported in Table S2, are in good agreement with available theoretical and experimental data.\cite{merikoski1997effect,merikoski1995diffusion,perkins1993self,boisvert1997self,liu1991eam,scheffler1997physical,durr1995island,ernst1992nucleation, Hansen1991Selfdiffusion,montalenti1999jumps} This last test validates the ability of the potential to correctly reproduce the dynamic properties of the surface atoms.

We used such \textit{NN}-potential to simulate large FCC(111), (211), (110), and (210) surface models composed of $2400$ Cu atoms ($2304$ for (210)) \textit{via} deep potential molecular dynamics (DPMD) simulations. All Cu surface models have depth $>15$\AA{} and replicate on the \textit{xy} plane through periodic boundary conditions, effectively modeling a portion of the bulk of infinite surfaces (Figure~\ref{fig1}c). 
During the DPMD simulations, the 2 bottom layers in the surface models are kept fixed, while all other atoms are free to move.
Such setup and system sizes prevent finite-size effects and guarantee reliable modeling of the structural dynamics of these Cu surfaces (see Methods for details).

\begin{figure*}[ht!]
\centering
\includegraphics[width=\textwidth]{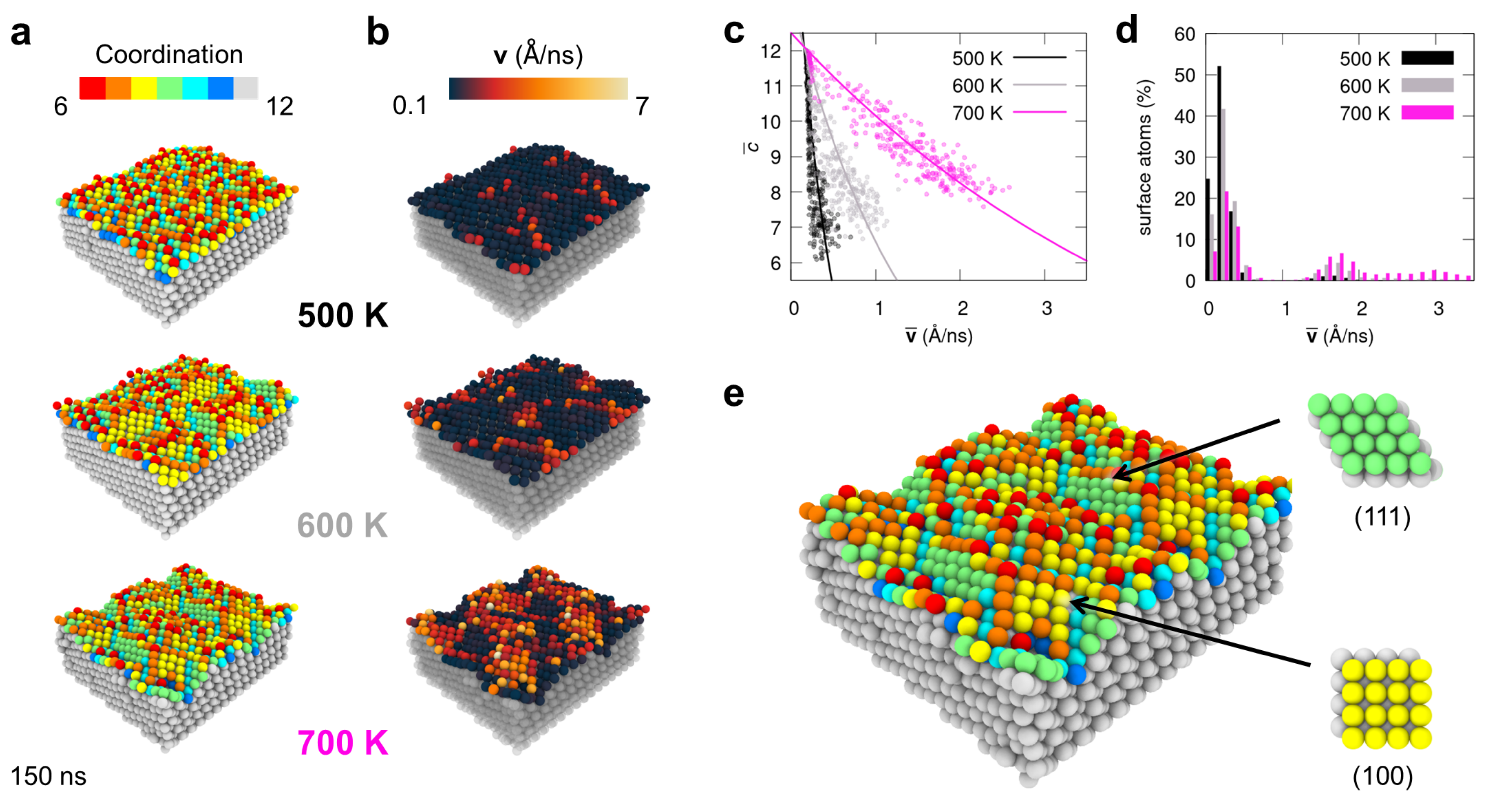}
\caption{\textbf{Dynamic diversity in a metal surface.}
\textbf{a} The Cu(210) surface after 150~ns of DPMD simulations at 500, 600 and 700 K of temperature (top-to-down). Atoms are colored based on their coordination.
\textbf{b} Relative diffusion velocity in the (210) surface at the different temperatures:  the Cu atoms are colored with the relative velocity, estimated for each atom in the system as the atomic displacement in the time interval of $\Delta t=1500$ ps ($\Delta r$/$\Delta t$, expressed in \AA{}~ns$^{-1}$). Dark \textit{vs.} light colors in the snapshots identify static \textit{vs.} dynamic regions on the Cu surfaces (bulk and sub-surface atoms are shown transparent for clarity). 
\textbf{c} Correlation between average coordination number and velocity for the atoms in the (210) surface at the three temperatures. Solid lines are exponential fits to guide the eye.
\textbf{d} Histograms of the atomic velocity distributions in the (210) surface at 500, 600 and 700 K.
\textbf{e} Zoomed detail of the coordination analysis for the (210) surface at 700 K, highlighting the emergence of (100) and (111) domains along the DPMD simulation (yellow and green details).
}
\label{fig2}
\end{figure*}

All systems underwent 150 ns of DPMD simulations at 500, 600 and 700~K, the last 75 ns of the simulations being representative of the equilibrium of the surfaces (see Supplementary Figures~S2~S3 and~S4 and Methods section). 
Changes in the atoms' colors in Figure~\ref{fig1}d indicate changes in atomic coordination, atomic movements and reconfigurations along the DPMD simulations.
The DPMD simulations show how the various surfaces have different dynamic behaviors: (111) is substantially persistent/static, while (211) and (110) surfaces are much more dynamic. The (210) surface is found the most dynamic of the studied surfaces.
These results are in agreement with the order of stability determined by the corresponding surfaces energies, and more importantly with experimental evidences. For example, experimental studies proved that the (111) surface does not undergo any structural modification or premelting below the melting temperature, but the surface enters a disordered state only at the melting temperature.\cite{Chae1996111,Alrawi2002111} 

For what concerns the (110) surface, a large body of experimental evidences show that this orientation is characterized by an increasing disorder induced by thermal energy already at T > 550 K. This degree of disorder could not be simply assigned to thermal anharmonicity because it was much larger and for this reason was named "enhanced anharmonicity"\cite{Yang1991Enhanced}. Tosatti and coworkers attributed this phenomenon to a missing-row type reconstruction, given the small difference between the surface energies of the reconstructed and unreconstructed surfaces\cite{Trayanov1989missingrowrough,trayanov1990anisotropic}. Our simulations confirm this interpretation as can be observed in Figure~S5. Here we compare the snapshot of our Cu (110) surface after 150 ns at 700 K with the corresponding ideal (110)(1$\times$2) and (1$\times$3) reconstructions coloring the atoms according to their coordination. It is immediate to notice the formation of a (1$\times$3) missing-row type reconstruction in the section highlighted by the dashed box. 

More recently, also the Cu (211) has been the object of several studies, given its reactivity toward the methanol synthesis.\cite{Science2012activesitemethanol}. Witte et al\cite{Witte1998Oinduced211}. have found that upon oxygen adsorption already below room temperature also the (211) surface undergoes reconstruction forming a (211)(2$\times$1) surface. The same type of reconstruction is observed in our simulations at 600 and 700 K. In figure S6 we compare the ideal (211)(2x1) reconstructed surface with a snapshot of the (211) surface after 150 ns at 700 K. As for the (110), also in this case it easily detectable a reconstruction of the surface toward the (211)(2$\times$1) surface.  

From a dynamic perspective the (210) surface isthe most interesting as it is found dynamic at all temperatures, with its dynamics increasing with temperature (Figure~\ref{fig2}a).
Deeper microscopic analyses reveal a considerable dynamic diversity in the behaviors of the atoms in the (210) surface. 
While the temperature is set globally in these DPMD simulation (see Methods for details), the data show that there are atoms that move faster and atoms that are more static on the surface. This can be inferred \textit{via} estimating along the DPMD the displacement of the individual atoms in the surface in the time interval $\Delta t=1500$ ps ($\Delta r$/$\Delta t$).

For example, bright atoms in Figure~\ref{fig2}b move by tens of nanometers during the DPMD simulations, while black atoms just vibrate around their lattice position.
 
While in these simulations the atomistic surface models are thermostated and on average their temperature plateau to 500, 600 and 700 K, these analyses demonstrate that on a nanosecond timescale window, the structure and dynamics on the surfaces is not uniform. In particular, the  plots of Figure~\ref{fig2}b show that these surfaces are dynamically diverse, being populated by domains that are more static and others that are more dynamic.
Still, it is noteworthy to add that all these domains are continuously destroyed and formed with different time-scales, and are also in continuous dynamic exchange with each other as it will be characterized in detail in Figures~\ref{fig3},~\ref{fig5}, and~\ref{fig6}.
The plots of Figure~\ref{fig2}c relate the average velocities and coordination numbers of the atoms. The minimum atomic coordination increases from $\sim 6$ (black data) to $\sim 8$ (in pink) while increasing the temperature in the (210) surface. 
The histograms of the atomic velocity distributions (Figure~\ref{fig2}d) indicate the variability with which the (210) surface atoms move, a few having high relative mobility while the majority of them are more static.
Figure~\ref{fig2}a shows that at 500 K (top) the ideal structure of the Cu (210) surface is better preserved during the DPMD than at 700 K (bottom), where the increased kinetic/thermal energy triggers more considerable disordering and faceting which in turn produces a surface configuration populated by more stable atomic arrangements (facets) with increased coordination which corresponds to (111) facets (in green) and (100) facets (in yellow).
A zoom onto the (210) surface at 700 K (Figure~\ref{fig2}e) shows how such green domains correspond to (111) islands, non-native of this surface. Yellow colored (100) square domains, non-native of (210), are also visible. It is interesting to notice that similar results were observed in the experimental work of Kirby et al.\cite{Kirby1980faceting210} In particular, they detected a faceting phenomenon of the Cu(210) surface induced by activated nitrogen already at room temperature.
Another interesting aspect revealed by our analysis comes from the comparison of the colored domains in Figures \ref{fig2}e \textit{vs.} \ref{fig2}b, which provides an insight into the dynamic diversity of the surface, revealing how non-native (111) domains (Figure~\ref{fig2}e: green) are more static while (100) ones (in yellow) correspond to more dynamic regions in Figure~\ref{fig2}b. 
Noteworthy, along the DPMD simulations the atoms change color dynamically in these Cu surfaces, demonstrating the dynamic equilibrium present within them (see Supplementary Movies S1 and S2).

\subsection{Unsupervised machine learning of structure and dynamics of a copper surface}

To obtain a more robust and general quantitative analysis, we turned to an advanced data-driven approach recently proven useful for reconstructing the structural/dynamical complexity of various types of molecular systems.\cite{gardin2022classifying,bian2021electrostatic,gasparotto2019identifying,capelli2022ephemeral,capelli2021data} 
We use SOAP vectors as a high-dimensional descriptors of the local environments surrounding each atom in these surfaces. 
Calculation of the SOAP spectra of all atoms along the DPMD simulations allows (i) to classify the local atomic environments that populate/emerge within the Cu surface in equilibrium conditions based on their levels of order/disorder and similarity, and (ii) to reconstruct the entire Cu surface dynamics (see Methods for details).\cite{gardin2022classifying,bian2021electrostatic,capelli2021data,gasparotto2019identifying}
Key advantages of such analysis are that the SOAP descriptor is agnostic\cite{SOAPBartok} and the analysis is unsupervised and data-driven: \textit{i.e.}, it does not require prior knowledge of the systems, while the SOAP detected environments emerge directly from the DPMD trajectories (\textit{bottom-up} analysis).\cite{gardin2022classifying,bian2021electrostatic,capelli2021data,gasparotto2019identifying}

\begin{figure*}[ht!]
\centering
\includegraphics[width=0.84\textwidth]{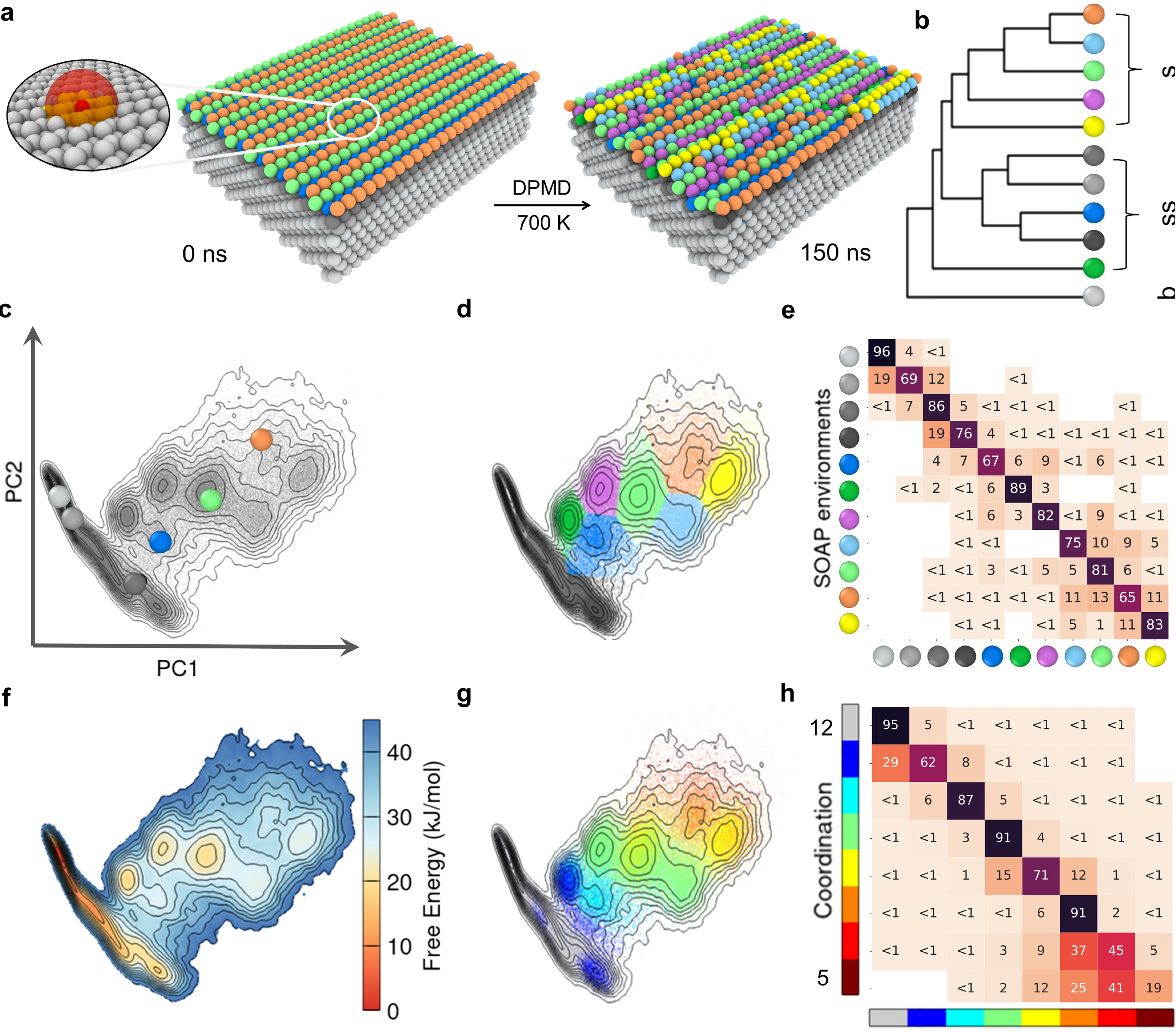}
\caption{\textbf{ML of atomic environments in the Cu(211) surface and of their dynamics.}
\textbf{a} Cu atoms on the (211) surface colored based on the SOAP environments emerging along the equilibrium (last 75 ns) DPMD simulation at 700 K.
\textbf{b} Hierarchical dendrogram connecting the 11 detected SOAP clusters, corresponding to 6 internal atomic environments -- 5 sub-surface (ss) and 1 bulk (b) -- plus 5 surface environments (s).   
\textbf{c} Projection on the first two principal components of the SOAP data PCA with density isolines. The colored dots indicate the position on the PCA of the native SOAP states present in the ideal (211) surface at the DPMD simulation start (\textbf{a},left: at 0 K). 
\textbf{d} Unsupervised clustering (HDBSCAN*) of SOAP data identifies 11 main SOAP environments (microstates) that emerge at the equilibrium in the (211) surface at 700 K. 
\textbf{e} Transition matrix reporting the normalized probabilities (in $\%$) for atoms to undergo transition between the SOAP clusters in a sampling time interval of $dt = 300$ ps.
\textbf{f} Free Energy Surface (FES) computed from the PCA of the SOAP density data (\textbf{c}).
\textbf{g} PCA projection of the SOAP dataset colored based on atomic coordination.
\textbf{f} Transition matrix reporting the normalized probabilities (in $\%$) for atomic transitions between the various coordination states along the DPMD ($dt = 300$ ps).
}
\label{fig3}

\end{figure*}

We perform our SOAP analysis on the DPMD trajectories of the various simulated Cu surfaces (see Supplementary Figure~S12). After 75~ns of DPMD, all surfaces reach a microscopic equilibrium where the atomic environments populating the systems do not change anymore (see Supplementary Figures~S2,S3,S4). The last 75~ns of DPMD, representative of the equilibrium of the modeled surfaces, are thus retained for the analyses.
From these, 250 snapshots -- one every $\Delta t= 300$ps -- are extracted and analyzed. In particular, we calculate the SOAP spectra of each of the topmost $1300$ atoms in the Cu surface, obtaining at each DPMD sampled snapshot $1300$ SOAP spectra representative of the pristine arrangements/dynamics of the atomic surfaces in the studied regimes, and a dataset of 325'000 SOAP spectra in total for each simulated case (see Methods section).
Unsupervised clustering of the SOAP data \textit{via} the Hierarchical Density-Based Spatial Clustering of Applications with Noise* (HDBSCAN*) algorithm~\cite{HDBSCANCampello2013,HDBSCANStarCampello2015} identifies the main SOAP clusters (states/environments) populating the equilibrium DPMD trajectories (Figure~\ref{fig3}a,b). 
As an example, HDBSCAN* identifies 11 main SOAP clusters in the Cu(211) surface at 700 K (Figure~\ref{fig3}a,b). These correspond to the density peaks seen in the PCA of the SOAP data in Figure~\ref{fig3}c,d, namely, to the most ``visited'' atomic environments during the DPMD simulation.
The dendrogram of Figure~\ref{fig3}b shows the adjacency between the detected SOAP micro-clusters (based on their similarity), revealing three main macro-groups: surface (s), sub-surface and bulk (b) atomic environments. Shades of gray in Figures~\ref{fig3}a,b correspond to deeper sub-surface layers connected to the bulk (light gray). Dark blue and green identify states at the interface between sub-surface and surface.
Brighter colors identify the different surface states (s). 
The colored dots on the PCA projection of Figure~\ref{fig3}c indicate which SOAP states are present in the ideal (211) surface at 0 K (at DPMD start). 
Comparing Figure~\ref{fig3}c and Figure~\ref{fig3}d it is clear how, in terms of external surface states, only the orange, light-green and blue SOAP environments are native of the ideal (211) surface (at 0 K). All other surface environments that emerge along the DPMD at 700 K (yellow, cyan, purple, etc.) are non-native states, which emerge with temperature.

Since the detected SOAP environments are well-sampled along the DPMD, we know the clusters density at the equilibrium, and we have information on the SOAP environment each atom belongs to at every sampled DPMD snapshot, we can reconstruct the dynamics and thermodynamics of the Cu surface.
The transition matrix of Figure~\ref{fig3}e reports the normalized probabilities for an atom belonging to a given SOAP environment at a time $t$ to remain in that environment (diagonal entries) or to undergo transition to a different SOAP environment (off-diagonal entries) at $t+\Delta t$ (\textit{i.e.}, after $\Delta t = 300$ ps in this analysis) in the Cu(211) surface at 700 K. Such transition matrices are non-symmetric as they are normalized to have the rows summing to 100, while the raw non-normalized matrices are conversely symmetric, as the Cu surface is at the equilibrium (see Supplementary Figure~S10).
In general, the higher are the numbers on the diagonal of the matrix, the more persistent is the surface. \textit{Vice versa}, the higher are the off-diagonal probabilities, the more probable are the atomic transition between the SOAP states in $\Delta t$ and the more dynamic is thus the surface.
The fact that the transition matrix of Figure~\ref{fig3}e is rich of off-diagonal entries demonstrates the rich dynamics present in this surface.
From such transition probabilities, one can estimate the characteristic transition rates/frequencies (\textit{i.e.}, by dividing the off-diagonal entries by $100 \times \Delta t$, being $\Delta t = 300$ ps in the analyses reported herein).
For example, in the (211) surface at 700 K, the atoms belonging to the purple SOAP environments undergo transition to blue with a probability of $\sim6\%$ every 300~ps, which corresponds to a transition rate of $\sim0.2~\mathrm{ns}^{-1}$ and transitions occurring in the timescale of tens of nanoseconds (assuming a single-step transition).
A consistent dynamics is obtained even changing the temporal resolution of the analysis ($\Delta t$), which proves the robustness of the obtained results (see Supplementary Figure~S11).\cite{Zuckermann2010transmat,Makarov2015transmat,Gupta2011transmat}
All entries $<1\%$ should be taken as qualitative, as they refer to events observed only sparsely along the DPMD.

This analysis shows there is no direct communication/exchange between bulk and surface states, while these may communicate only through intermediate sub-surface states. Even at 700~K, where the (211) surface is considerably dynamic, the transitions occur on a nanoseconds timescale. This demonstrates how the (211) surface has a "discrete" dynamics.

A similar "discrete" dynamics is observed for the (110) surface (Supplementary Figures~S12a-e). Also here the atoms of the first layers move via jumps across different crystallographic positions. This type of motion is correlated to the fact that when computing the g(r) of the surface the peak positions remain unchanged, however the surface atoms motion leads to a reduction of the peak intensities and a more diffuse background, as observed in Figure~S9. This type of atomic motion could explain the anomalous reduction of the intensity peaks of Cu (110) at T>550 K obtained by diffraction experiments.\cite{durr1991anomalous,trayanov1990anisotropic,zeybek2006thermal} This phenomenon was referred to as "enhanced anharmonicity" since the only corresponding property was an enhanced mean square displacement of the surface atoms. However, here we reinterpret this "enhanced anharmonicity" suggesting that its origin comes from the frequent jump-motion of the surface atoms.

The same analysis for the other surfaces studied herein shows that (210) is the most dynamic surface, with more fluid-like dynamics (Supplementary Figures~S12f-l). This behavior is also mirrored by a quasi-liquid like g(r) at 700 K (see Figure~S9). On the other hand, (111) surface is more static: only sparse transitions can be observed even at 700 K .

From the inverse exponential of the PCA density, it is also possible to estimate a pseudo Free Energy Surface (FES) for the Cu(211) surface model at 700~K (Figure~\ref{fig3}f). Such FES shows how at 700~K all surface SOAP states are separated from each other by relatively low free energy barriers within $\sim10-15 \mathrm{~kJ ~mol^{-1}}$ ($10 \mathrm{~kJ ~mol^{-1}}$ corresponds to $\sim2$~kT at 700~K). Their transitions can be thus efficiently sampled during an equilibrium DPMD (see Supplementary Movie S3). The SOAP analysis of Figures~\ref{fig3}d-f reveals how -- since the SOAP states have characteristic lifetimes and transition rates, and are in continuous interchange with each other -- at 700~K the (211) surface has just an average configuration that is purely statistical.
The great flexibility and the agnostic nature of such data-driven analysis come with the disadvantage of a non-straightforward interpretation.
There is no-straightforward correspondence between the detected SOAP states and their physical differences. Figure~\ref{fig3}g shows the same SOAP PCA colored according to the coordination number of each atom.
This shows how the SOAP analysis captures very well differences in coordination number between the atomic environments in the Cu surface, while at the same time a simple coordination analysis is less sensitive -- see, \textit{e.g.}, green domains in the PCA having the same coordination number ($\sim9$) but corresponding to different SOAP density peaks (Figure~\ref{fig3}g). 

From a broader perspective, this data-driven analysis shows that such Cu surfaces possess a non-trivial structural/dynamical complexity well below melting.
At the same time, these results underline the importance of relying on a structurally and dynamically accurate force field (as the \textit{NN}-potential used herein) to obtain meaningful insights on such a complexity.
The fact that new states, non-native/present in a given (ideal) surface may appear at finite temperature along the DPMD simulations poses fundamental questions on the elusive identity of these surfaces.
For example, are the new environments that emerge with a temperature closer to native environments present in the ideal surface (at 0 K), or, \textit{e.g.}, to other ones, native of different types of surfaces? To answer such questions we designed another complementary (\textit{top-down}) analysis.

\subsection{A dictionary of SOAP atomic environments \label{sub:dict}} 

\begin{figure*}[ht!]
\includegraphics[width=0.9\textwidth]{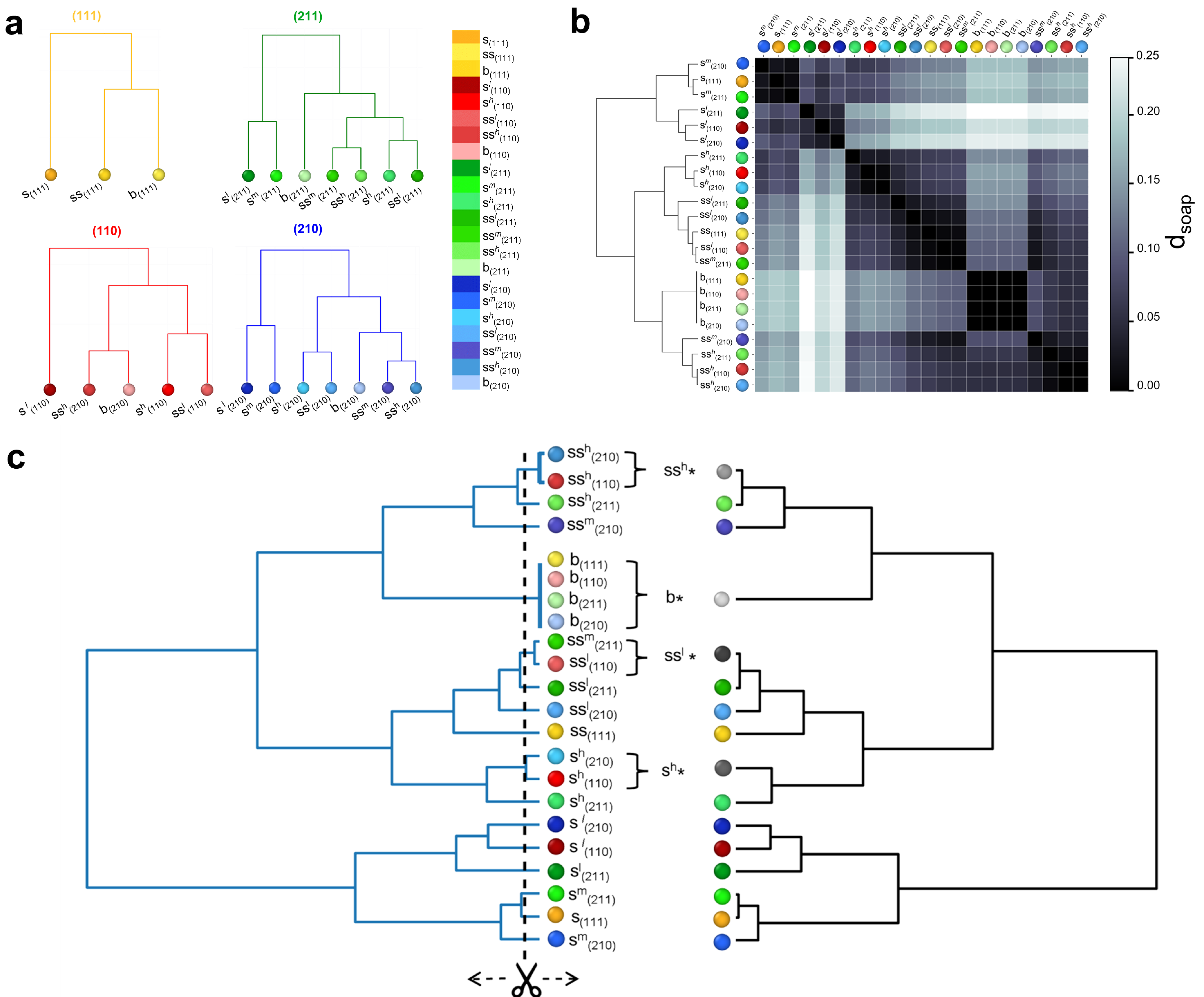}
\caption{\textbf{A SOAP dictionary for classifying atomic environments in Cu surfaces.} \textbf{a} Hierarchical dendrograms for the SOAP environments detected in the different ideal Cu surfaces (at 0 K): Cu(111) environments are shown in yellow, (211) in green, (110) in red, and (210) in blue. 
\textbf{b} Distance matrix: the color scale indicates the distance in the high-dimensional SOAP feature space ($d_{soap}$) between all SOAP environments in the Cu surfaces.
\textbf{c} Left: Hierarchical dendrogram (in blue) showing the similarity between all SOAP environments (also reported on the left of the $d_{soap}$ matrix, \textbf{b}). Right: same dendrogram cut at $d_{soap} \geq 0.01 $ (in black). This results in grouping very similar SOAP environments into common macro-clusters: \textit{e.g.}, bulk (b*), sub-surface (ss*), surface (s*).}
\label{fig4}
\end{figure*}

We created a  dictionary of SOAP atomic environments by computing the SOAP spectra for all atoms in the optimized  Cu(111), (110), (211), and (210) surfaces (at 0 K).
The analysis identifies 3 SOAP atomic environments in the ideal (111) surface -- Figure~\ref{fig4}a (yellow): one bulk (b$_{(111)}$), one sub-surface (ss$_{(111)}$) and one surface environment (s$_{(111)}$). 
The other surfaces are more structurally diverse: the ideal (110) is characterized by 5 main SOAP atomic environments (in red), while the ideal (210) and (211) are characterized by 7 SOAP environments (Figure~\ref{fig4}a: blue and green respectively). 
In total, we obtain 22 distinct SOAP spectra characteristic of the bulk, surface and sub-surface atomic environments proper of these ideal surfaces. 
We created a unique SOAP data-set containing all these SOAP environments and computed from their characteristic SOAP power spectra their mutual distances ($d_{soap}$) in the global high-dimensional SOAP feature space (see Methods for details).\cite{gardin2022classifying,capelli2021data,capelli2022ephemeral}  
Such $d_{soap}$ metrics allows quantifying the similarity between the various characteristic SOAP spectra,
providing a rich data-driven classification of all detected atomic environments present in these ideal surfaces.
The result is the distance matrix of Figure~\ref{fig4}b.
The colors of the matrix cells represent the SOAP distance ($d_{soap}$) between the various SOAP atomic environments: dark colors indicate very similar environments ($d_{soap} \sim0$), light colors identify structurally different atomic environments.
The dendrogram adjacent to the matrix (Figure~\ref{fig4}b: left) shows the hierarchical clustering of the various SOAP environments based on their similarity.
The matrix reveals dark macro-areas indicating SOAP environments that are nearly identical in the various surfaces --  \textit{i.e.}, bulk (b) environments, some high-coordination sub-surface ones (ss), etc.
In general, deep atomic environments are found quite similar in the various ideal surfaces.
The matrix also reveals non-obvious similarities between the high, medium, and low coordination surface environments: \textit{e.g.}, s$_{(111)}$ \textit{vs.} s$^{\text{m}}_{(211)}$ and s$^{\text{m}}_{(210)}$, or s$^{\text{l}}_{(110)}$ \textit{vs.} s$^{\text{l}}_{(210)}$. 

Figure~\ref{fig4}c shows how the complete dendrogram (left, in blue) can be cut in order to consider only detected SOAP distances greater than a minimum value, offering the opportunity of a variable resolution in the analysis.
For example, Figure~\ref{fig4}c (right) shows what the dendrogram becomes when considering only $d_{soap} \geq 0.01 $ (in black).
At this level of resolution, the bulk environments of all surfaces are grouped in a single bulk state (b*). The same happens for other very similar sub-surface (ss*) and surface (s*) environments.
While complete information is encoded in the pristine dendrogram, this offers the opportunity to modulate the noise/relevance trade-off of the analysis, focusing on differences that are really meaningful (\textit{e.g.}, distinguishing between the bulk states of these surfaces is useless, as these are identical SOAP environments).
As it will be demonstrated in the next section, this is important, for example, when using such SOAP data and $d_{soap}$ metric to track the similarity between the atomic environments emerging in the metal surface in equilibrium condition and those included in the SOAP dictionary (\textit{top-down} classification).

\subsection{Dynamic reconstructions and statistical equivalent identities of copper surfaces}

\begin{figure*}[ht!]
\centering
\includegraphics[width=0.92\textwidth]{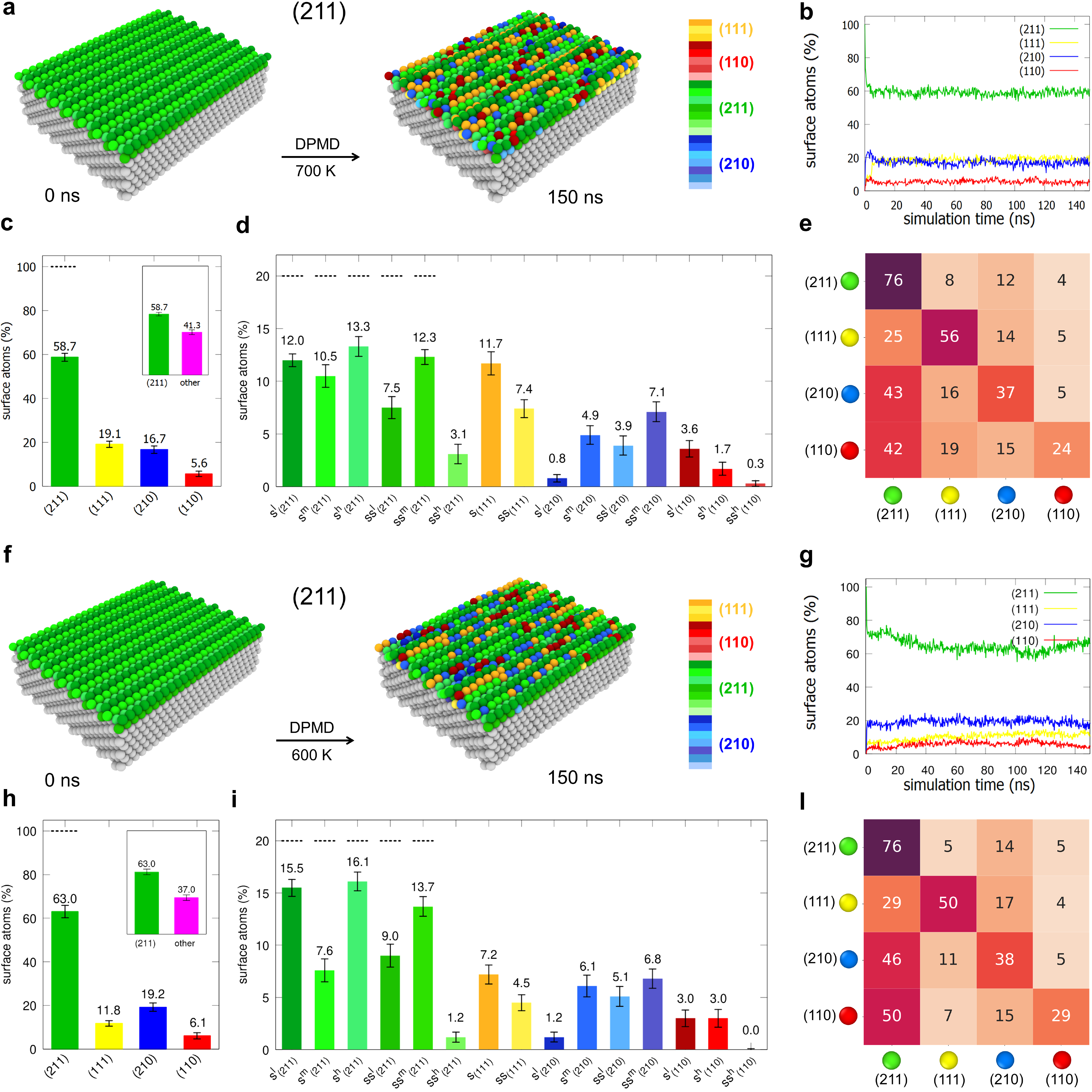}
\caption{\textbf{Dynamic reconstructions and equivalent identity of a Cu(211) surface.}
\textbf{a} Cu(211) at 0~ns (left) and after 150~ns of DPMD at 700~K (right): SOAP environments native of the ideal (211) surface are colored in green. Red, blue and yellow colors identify non-native atomic domains, proper of (110), (210) and (111) surfaces.
\textbf{b} Populations of the native and non-native environments (in $\%$) in the (211) surface at 700~K as a function of DPMD time. 
\textbf{c} Equilibrium composition of (211) at 700 K ($\%$ and standard deviations) in terms of native (green) and non-native domains (yellow, blue, red, combined in pink in the inset).
\textbf{d} Breakdown of (211) composition at 700 K. Dashed lines indicate the composition at DPMD start.
\textbf{e} Transition matrix showing the probabilities for atomic transitions in (211) between native and non-native environments at 700 K (within $\Delta t = 300$ ps).
\textbf{f-l} Same analyses for the (211) surface at 600 K.}
\label{fig5}

\end{figure*}

Starting from the Cu(211) surface at 700 K, at each timestep along the DPMD simulation we measure the $d_{soap}$ distance between the SOAP spectrum of each atom and all SOAP spectra characteristic environments present in the dictionary of Figure~\ref{fig4}.
At each DPMD timestep, each atom is then attributed to the closest SOAP environment/class (smallest $d_{soap}$) in the SOAP dictionary. 
This allows us to track the transformations in the surface along the DPMD and to estimate the reconstruction of non-native domains, their lifetime and dynamics in terms of atomic transitions between them.

Since we are interested in surface reconstructions and dynamics, in this phase we focus only on the five top-most layers of the simulated surfaces (see Methods).
Figure~\ref{fig5}a shows how the (211), ideally composed only of green native SOAP sites at 0 K (left), convert into local non-native domains, proper of (111), (110) and (210) ideal surfaces (right: yellow, red and blue respectively), along 150 ns of DPMD at 700 K. 
The environment populations of Figure~\ref{fig5}b show that the surface reaches a microscopic equilibrium along the DPMD, being populated of native and non-native domains. Along the DPMD, $\sim40 \%$ of native domains disappear converting into non-native domains (Figures~\ref{fig5}b,c).
On a statistical level, the (211) surface thus preserves its own identity only by $\sim60 \%$ at 700 K (Figure~\ref{fig5}c: inset).
In particular, $\sim19 \%$ of the emerging domains correspond to (111) environments, $\sim17 \%$ to (210), and $\sim6 \%$ to (110) ones (Figure~\ref{fig5}c).
Such surface reconfiguration is rather fast in this system (Figure~\ref{fig5}b).
Figure~\ref{fig5}d shows a breakdown of the detailed native and non-native environments that populate the surface in equilibrium conditions.
Among all the emerging non-native environments, s$_{(111)}$ is the predominant one ($\sim12 \%$), followed by ss$_{(111)}$ and ss$^{\text{m}}_{(210)}$, both constituting $\sim7 \%$ of the (211) surface at 700 K. 
While such analysis provides detailed quantitative information on the composition, structural diversity, and reconfiguration in the metal surface, it is interesting to interrogate on the dynamical features of such phenomena.

The transition matrix of Figure~\ref{fig5}e reports the normalized probabilities for the atom transitions between native and non-native environments in the (211) surface at 700 K (in $\Delta t = 300$ ps). 
The diagonal entries show that only the native (green) and yellow (111) domains have a residence probability $>50 \%$. This means that at the temporal resolution of this analysis, these are somewhat persistent domains. On the contrary, (210) and (110) atomic environments have persistence probabilities well $<50 \%$. Such domains are considerably more dynamic, and the atoms composing them have a higher probability to re-convert into native domains (red-to-green and blue-to-green transition probabilities $\sim 42-43 \%$) than to remain there in $\Delta t = 300$ ps.
This provides a picture reminiscent to what seen in Figure~\ref{fig2} for the (210) surface, where above the H\"uttig temperature such dynamically diverse surface appears as composed of stable domains coexisting in equilibrium with more dynamic ones (Supplementary Movie S4).

Decreasing the temperature to 600 K, the behavior of (211) does not change substantially (Figures~\ref{fig5}f-l). 
In general, the transformation of the (211) surface is rather similar, while it occurs slightly slower at 600 K than at 700 K (Figure~\ref{fig5}g).
The appearance of (111) domains is just slightly reduced than at 700 K (Figures~\ref{fig5}h,i: $\sim12 \%$ at 600 K \textit{vs.} $\sim19 \%$ at 700 K). 
The transition matrix of Figure~\ref{fig5}l is also very similar to that of Figure~\ref{fig5}e.
This demonstrates how the phenomena occurring in these regimes are thermodynamically driven, being, \textit{e.g.}, the surface energy of (111) lower than that of (211) (See Table S1).
Conversely, the same analysis demonstrated how the (211) surface appears as substantially static at 500 K (H\"uttig temperature of Cu: 447 K) in the same DPMD timescales. In such a regime the thermal bath is evidently insufficient to trigger the reconfiguration and the surface remains trapped in the ideal (211) configuration (Supplementary Figure~S13).
In section A of the Results, it was highlighted the similarity between the final state of the simulated Cu (211) surface and the corresponding missing-row (211)(2$\times$1) reconstruction. We deemed thus interesting to analyze with our SOAP based approach the ideal (211)(2$\times$1) reconstructed surface, and noticeably this surface shows a number of yellow (111) atomic domains which is in line with what has been observed for the Cu (211) surface simulated at 700 K (See Figure~S8). This confirms the ability of the present approach to detect (111) facets in reconstructed surfaces. The differences between the simulated and reconstructed (211) surfaces originate from the actual motion of the atoms which undergo a frequent changes in their surrounding environment.

\begin{figure*}[ht!]
\centering
\includegraphics[width=0.92\textwidth]{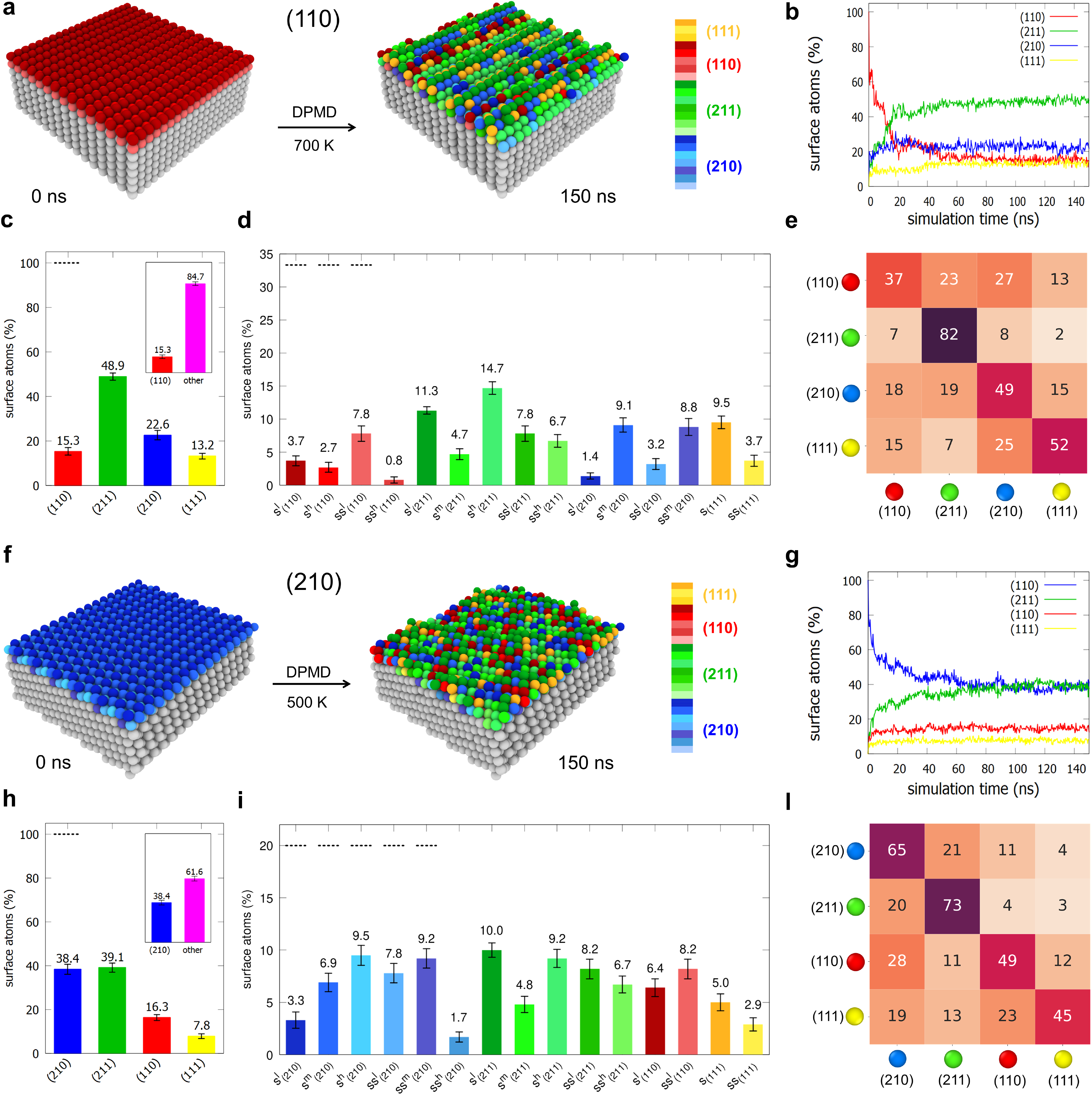}
\caption{\textbf{Dynamic reconstructions and equivalent identity of other Cu surfaces.}
\textbf{a} Cu(110) at 0~ns (left) and after 150~ns of DPMD at 700~K (right): SOAP environments native of ideal (110) are colored in red. Green, blue and yellow colors identify non-native environments proper of (211), (210) and (111) surfaces.
\textbf{b} Environment populations in the (110) surface at 700~K as a function of DPMD time. 
\textbf{c} Equilibrium composition of (110) at 700 K ($\%$ and standard deviations). Inset: native domains in red, non-native in pink.
\textbf{d} Breakdown of (110) composition at 700 K. Dashed lines indicate the composition at DPMD start.
\textbf{e} Transition matrix showing the probabilities for atomic transitions in (110) at 700 K (within $\Delta t = 300$ ps).
\textbf{f-l} Same analyses for the (210) surface at 500 K.
Same analyses for surfaces at other temperatures are reported in the Supplementary Information.}
\label{fig6}

\end{figure*}

Comparing the behavior of the other Cu surfaces, the same analysis reveals that the (110) surface is highly dynamic and substantially unstable at 700 K (Figures~\ref{fig6}a-e). The snapshots of Figure~\ref{fig6}a show how during 150 ns of DPMD, the surface becomes largely populated of non-red colors, and mainly of green (211) domains (Supplementary Movie S5).
Figure~\ref{fig6}b indicates a substantial reconfiguration of the (110) surface. 
The instability of (110) at 700 K is manifest in the fact that the surface is reconstructed by $\sim85 \%$.
The native red environments drop to $<20 \%$ of the surface, reconstructing in large part (211) green domains (Figure~\ref{fig6}c: rising to $\sim50\%$).
Starting from an ideal (110) configuration, such surface evolves towards reconstructing a different, more stable surface. Similar to what seen for (210) at 700 K, the transition matrix of Figure~\ref{fig6}e reveals how the residual red native domains are also highly dynamic (survival probability $<40 \%$), which fits well with their relatively high surface energy,\cite{Jian_Min_2004,WANG2014surfene} while the persistent domains in this surface are non-native environments.
As seen experimentally for other metals,\cite{Titmuss1996reconstruction,Koch1992reconstruction,KOCH2000reconstruction} also the Cu(110) surface appears unstable at 700 K and reconstructs large surface domains structurally/dynamically similar to the (211) surface ones (Figure~\ref{fig5}).
Also at 600 K, the (110) surface has a dynamics similar to that of (211) at the same temperature, while the reconstruction of (211) domains is much slower than at 700 K.
Like for (211), at 500 K also the (110) surface is substantially immobile and preserves its identity: the thermal bath is insufficient to trigger the reconstruction in the timescales accessible \textit{via} these DPMD simulations (see Supplementary Figure~S14).

The emergence of a large number of green (211) atomic domains can be quite puzzling. However, it can be explained through a SOAP analysis of the missing-row (110)(1$\times$3) reconstructed surface. When coloring this surface with our SOAP based dictionary we observe that the atoms of the surface are composed mainly of green (211) and yellow (111) atomic domains. This evidence shows that the atomic environments that are native of the (110)(1$\times$3) reconstructed surface are actually very close in terms of soap distance to those of the (211) surface (See Figure~S7). 

The Cu(210) surface is way more dynamic than both (211) and (110) surfaces.
Among all studied surfaces, (210) is that having the least coordinated atoms: the s$^{\text{l}}_{(210)}$ are the only atoms with coordination 6 in the whole dataset (see Figure~\ref{fig1}c).
Figures~\ref{fig6}f-l show the analysis for (210) at 500 K. Even so close to the H\"uttig temperature, this surface undergoes considerable reconstruction (see Supplementary Movie S6). 
The (210) reconstructs non-native domains by $>60 \%$, preserving its identity only by $<40 \%$ (Figures~\ref{fig6}g,h). 
Increasing the temperature to 600 K or 700 K has the unique effect of accelerating such reconfiguration, while the equilibrium populations remain substantially preserved (Supplementary Figure~S15). 
This fits well with the higher energy of this Cu surface.\cite{Jian_Min_2004,WANG2014surfene}
One difference is in the dynamics of such atomic environments. The transition matrix of Figure~\ref{fig6}l shows diagonal entries very close or higher than $50 \%$. At the resolution of our analysis, the dynamics that emerges in this surface at 500 K is discrete (solid dynamics). 
Conversely, increasing the temperature to 700 K creates dynamically-persistent solid-like domains -- native blue (210) and non-native green (211) domains -- coexisting with dynamic domains (Supplementary Figure~S15).
This confirms that, also in this case, increasing the temperature does not generate a uniform increase of the dynamics of atoms, but the emergence of local dynamic domains and a non-uniform dynamically-diverse surface. 

The last case that we compare is the close-packed Cu(111) surface. Even the simple coordination analysis of Figure~\ref{fig1} clearly shows that this surface is very stable\cite{Jian_Min_2004,WANG2014surfene} and does not undergo any considerable reconstructions in such regimes. Even at 700 K surface atoms with coordination $\neq $ 9 emerge only sparsely and as statistical local fluctuations, indicating vibrations rather than reconstructions.

\section{Discussion}
We report a data-driven approach that allows resolving at atomistic-resolution the complex structural dynamics of metal surfaces above the H\"uttig temperature. As a case study, we use Cu surfaces. However, the approach is versatile and can be applied to other metal systems.
The approach provides a detailed microscopic characterization of the atomic environments composing such dynamically diverse surfaces, the rates with which these emerge/disappear, their residence time and persistence (see, \textit{e.g.}, Figure~\ref{fig3}). The development of a dictionary of SOAP atomic Cu surface environments (Figure~\ref{fig4}) allows for a data-driven analysis of the similarity/differences between the local motifs that appear in dynamic equilibrium conditions on the different Cu surfaces (Figures~\ref{fig5},\ref{fig6}c,d,h,i). This provides an exquisitely statistical picture of these metal surfaces, and a data-driven estimation of their "equivalent identity" in dynamic regimes (Figures~\ref{fig5},\ref{fig6}). Knowing what local environments emerge, how often, for how long is a prime requisite to understand what a metal surface looks like and its properties in determined thermodynamic conditions.

The developed \textit{NN}-potential allows dynamically-reliable DPMD simulations of relatively-large Cu surface models composed of $2400$ atoms (replicating on \textit{xy} through periodic boundary conditions).
Noteworthy, the transition matrices of Figures~\ref{fig5},\ref{fig6} show transition probabilities ranging $\sim5-40 \%$ in $\Delta t = 300$ ps, revealing a rich microscopic dynamics in such metal surfaces with characteristic times for the transitions between the various environments in the order of nanoseconds. 
On a technical standpoint, this shows how such simulations provide access to information extremely difficult to attain with other approaches. 

From a scientific point of view, metal surfaces in most cases are still studied treating the surface as a rigid object, however, the rich structural dynamics seen in these metallic materials at temperature-regimes of 500-700 K in our simulations indicate that the actual scenario is much more complex and that the intrinsic dynamics of the metal surface must be explicitly accounted to understand surface properties.
In particular, it is intriguing to note that the results of Figures~\ref{fig3},\ref{fig5},\ref{fig6} provide a picture of such metal surfaces that is quite far from that of hard materials, revealing internal dynamic equilibria and a structural/dynamical diversity that, in a sense, is reminiscent of that of soft dynamic materials.\cite{gasparotto2019identifying,gardin2022classifying, bochicchio2017into}

In perspective, our data-driven approach offers remarkable opportunities to relate the innate structural dynamics of metals to their properties. While here we are interested in resolving the complex structural dynamics of metal surfaces \textit{per se}, we envisage that this will have considerable impact in various fields, from the study of their mechanical properties to, \textit{e.g.}, heterogeneous catalysis,  where the dynamical emergence of local atomic environments with utterly different reactivity and survival lifetimes may have a strong impact on the catalytic activity.\cite{grosse2018dynamic,li2020electrochemically,higham2020mechanism,wang2004cluster,behrens2012active,Delgado-Callico2021,gazzarrini2021born} 
In general, this study reveals and resolves the complex dynamic character of metals in dynamic regimes, demonstrating how these cannot be simply studied as static structures even far from the melting temperature.
This changes the way we look at such materials, opening new exciting directions toward data-driven statistical reinterpretations of their properties.

\section{Methods}
\subsection{Neural network representation of the inter-atomic interaction potential}
The DFT database needed to train the Cu \textit{NN}-potential was generated extracting configurations along \textit{ab initio} molecular dynamics (AIMD) trajectories of small Cu systems. Simulations were performed using the PWscf code of Quantum ESPRESSO.\cite{giannozzi2009quantum} In these calculations, PBE exchange-correlation density functional\cite{perdew1996generalized} forces were used to propagate the nuclear dynamics. Preliminary tests demonstrated that this functional provides a reliable representation of the metal surface, free of empirical corrections, and constitutes an ideal compromise between accuracy and computational cost. Moreover, the same approach was also demonstrated reliable for other metals, e.g. for gold and silver surfaces.\cite{andolina-AuAgNP}
Ultrasoft RRKJ pseudopotentials\cite{rappe1990optimized} replaced explicit core-valence electron interactions, while electron density and wavefunctions were expanded in plane-waves with energy cutoffs of 220 and 50~Ry respectively. Occupation was treated by the cold smearing technique of Marzari\cite{marzari1999thermal} \textit{et al.} with a Gaussian spreading of 0.01~Ry. The Brillouin zone was sampled using a $2 \times 2 \times 2$ Monkhost–Pack k-point grid\cite{monkhorst1976special} for the bulk structure while a $2 \times 2 \times 1$ k-point grid was used for the slab models.  

Convergence against cutoff energy, Monkhost–Pack sampling, and occupation was tested and the setup described was chosen as the best compromise between feasibility and accuracy. 
The AIMD simulations were carried out with a time step of 1.0~fs in a constant volume and temperature (NVT) ensemble using the stochastic velocity rescaling thermostat of Bussi \textit{et al.}\cite{bussi2007canonical} In order to span a larger portion of the configurational space we simulated the systems at temperatures ranging between 500 and 700~K.

The AIMD simulated systems include both bulk and surface structures. The bulk was modeled by a periodically repeated super-cell of size $7.1638 \times 7.1638 \times 7.1638$~\AA{} containing 32 atoms. 
For surface calculations, 4 slabs models were used to construct the (100), (110), (111), (211) and (210) copper surfaces. Slabs with a different number of atomic layers were built and a vacuum layer was set in \textit{z} direction. The first two bottom Cu layers were kept fixed during optimization and AIMD simulations.

We trained the Cu \textit{NN}-potential using the DeePMD-kit package.\cite{deepMDPRL,deepMD} The smooth version of the deep potential model was adopted, with a cut-off radius of 6.0~\AA{}. To remove the discontinuity introduced by the cut-off, the 1/r term in the network construction is smoothly switched-off by a cosine shape function from 1.0 to 6.0~\AA{}. The filter (embedding) network has three layers with (25, 50, 100) nodes/layer and the fitting net is composed of three layers, with 240 nodes each. The network is trained with the ADAM optimizer,\cite{kingma2014adam} with an exponentially decaying learning rate from $1.0 \times 10^{-3}$ to $5.0 \times 10^{-8}$. 
The batch size was chosen as 4. The pre-factors of the energy and the force terms in the loss function change during the optimization process from 1 to 10 and from 1000 to 1, respectively. The final model used for the production run was trained for $10.0 \times 10^{6}$ steps (see Supplementary Figure~S1).

The choice of the training data-set is a crucial step in the training of an \textit{NN}-potential.
For this reason, we used the configurations collected along the DFT MD simulations to train a first "guess" Cu \textit{NN}-potential and we then used it to run several DPMD simulations for the different surfaces at different temperatures. This allowed to explore a larger portion of the configurational space accessible in the temperature range of interest (500-700~K) and to efficiently extract new configurations to add to the training data-set enriching further the \textit{NN}-potential.
This active learning protocol follows the same procedure introduced by Deringer and Cs\'anyi\cite{Deringer2017} and later implemented in the DeepMD software package by Zhang et al.\cite{dpgen} 
Here, the criterion used to select new configurations is based on the agreement on the forces predictions made by an ensemble of 4 \textit{NN}-potentials, which have been trained on the same reference data-set but with different initial weights.
We measured the model deviation as the maximum (over the force components) of the standard deviation on the forces predicted by such an ensemble of \textit{NN}-potentials.
Whenever the model deviation for one configuration was in the range of $[27-240 \times 10^{-3}]~eV/\text{\AA{}}$ the corresponding structure was included in the new training data-set.   
The iterative process continues until when no new configurations are visited, so that the \textit{NN}-potential can be considered ``complete'' for the sampled conditions. 
A total of 10'000 configurations were used for the training and the final root mean square errors of the \textit{NN}-potential on the testing set are equal to 1.0 meV/atom for the energy and 40 meV/~\AA{} for the force.

To verify that the use of $\sim100$ atom surface environment patches (accounting up to the $\sim4-5$th neighbor) for the training allows providing a \textit{NN}-potential that is robust and complete (\textit{e.g.}, free from spurious finite-size effects), we also performed additional tests using larger surface patches ($\sim600$ atoms) for the training. Our tests demonstrated that the deviations in the forces and energies thus estimated are systematically found within the training-testing errors, confirming thus the robustness of the approach adopted for the study of such systems and of the obtained \textit{NN}-potential.

Moreover, the fine sampling of atomic configurations and associated energies guaranteed by the DFT MD calculations allowed to include in the \textit{NN}-potential training data-set configurations corresponding both to local minima as well as to the transition states. In this way, the obtained \textit{NN}-potential is trained to represent with DFT accuracy the differences in energy between different sampled configurations, the transition barriers between them, and thus the transition kinetics.

This provided us with an atomistic \textit{NN}-potential with DFT precision in the treatment of the structure, energy and dynamics of Cu surfaces.
The \textit{NN}-potential is also rich in that, thanks to the inclusion of various types of surface environments in the training set (Figure~\ref{fig1}b), this is trained to cover a variety of atomic configurations -- from those, more stable, present in the FCC bulk and (111) surface at 500~K, to those emerging, \textit{e.g.}, in the Cu(210) surface, which is highly dynamic (and nearly pre-melted) at 700~K.
The use of a structurally and dynamically accurate Cu force field is key in our case, as it allows to reconstruct the internal atomic dynamics of Cu surfaces in an accurate and reliable way. Nonetheless, on a qualitative point of view, rather similar internal dynamics of the Cu surfaces has been obtained also in MD simulations using a different general-purpose and widely-used force field for copper available in the literature,\cite{CyrotDucastelle_bindingTransition,Gupta_energy,Rosato1989} confirming how the rich internal dynamics seen in our DMPD simulations is not exclusive of our \textit{NN}-potential and is somehow innate in these surfaces at this regimes.

\subsection{Atomistic surface models and DPMD Simulations}
All production simulations have been conducted using FCC Cu(111), (110), (210), (211) surface models composed of $2400$ atoms (only (210) has $2304$) replicating on the \textit{xy} plane through periodic boundary conditions, and having thickness $>15~$\AA{}.
During all simulations, the two bottom-most atomic layers were kept as fixed (these two layers are then not considered in all our analyses).
Preliminary tests demonstrated that such size is large enough to prevent finite-size effects on the structural dynamics of the surfaces with the adopted setup.

All DPMD simulations were performed using LAMMPS.~\cite{LAMMPS} The temperature was controlled using the stochastic velocity rescaling thermostat of Bussi et al.\cite{bussi2007canonical} The integration time step used is 1~fs and the relaxation time for the thermostat was set to 0.1~ps.
Coordination and velocity analyses of surface atoms were performed with the software OVITO.\cite{stukowski2009visualization}
Coordination was calculated as the number of neighbors within a certain cutoff ($R_{cutoff}=3.1$~\AA{}), that corresponds to the first minimum of the Cu-Cu radial distribution function. Surface atoms velocities in the analysis of Figure~\ref{fig2}b were calculated as $\Delta r/\Delta t$, where $\Delta r$ was computed as the atomic position displacement within a certain time window $\Delta t=1500$ ps. Both atomic coordination numbers and velocities reported in Figure\ref{fig2}c-d were averaged over the second half of the trajectory.    

\subsection{Smooth Overlap of Atomic Position}

The Smooth Overlap of Atomic Position (SOAP) is a many-body atomic descriptor, conceived to accurately reproduce particle densities of a given system of interest.\cite{SOAPBartok}
In our analyses, we center one SOAP vector in each atom of the surface models. For each atom at each timestep, this provides a SOAP "power spectrum", which is a high-dimensional descriptor of the level of order/disorder in the surrounding of each atom in the system.
The advantage of the SOAP descriptor is in that it is general and abstract, thus it can be used in a flexible way also, in principle, for other metals as well as for other types of molecular systems.\cite{gardin2022classifying,gasparotto2019identifying,capelli2021data,capelli2022ephemeral,bian2021electrostatic}

Given a system conformation $\Gamma$ in the 3D space, the local atomic density, $\rho_{i}(\Gamma,\vec{r})$, is defined in the neighborhood of every SOAP center within a spatial cutoff $r_{cut}$; 
the SOAP power spectrum can be evaluated projecting $\rho_{i}(\Gamma,\vec{r})$ onto a basis of orthogonal radial functions $g_{n}(r)$ and spherical harmonics $Y_{lm}(\theta,\phi)$, which for the $i$-th can be expressed as:
\begin{equation}
\rho_{i}(\Gamma,\vec{r})= \displaystyle\sum_{j \in r_{cut}} \displaystyle\sum_{nml} c^{j}_{nlm}(\Gamma)g_{n}(r)Y_{lm}(\theta,\phi)
\label{soap_density}
\end{equation}
where $c^{j}_{nlm}$ are the spherical harmonics and radial functions expansions coefficient, with that $j$ runs over all the sites within  $r_{cut}$.

The SOAP power spectrum vector can be analytically derived\cite{SOAPBartok} from eq.~\ref{soap_density} and can be written as:
\begin{equation}
p(\Gamma)_{nn^{'}l}=\pi\sqrt{\frac{8}{2l +1 }}\displaystyle\sum_{m=-l}^{l}c^{*}_{nlm}(\Gamma)c{n'lm}(\Gamma)
\label{soap_powerspectra}
\end{equation}

Once SOAP feature vectors have been calculated, the similarity between the various atomic environments can be inferred via a distance metric, as long as the dimensionality of the compared SOAP vectors is the same, as in our case, where we used for nmax,lmax=8 (found effective and sensitive enough in our case).

More precisely, we can define a measure of the similarity between two environments centered in two sites, building a linear polynomial kernel of their density representations; this can be simply reduced to the dot product of power spectra, defined in eq.~\ref{soap_powerspectra}. The SOAP distance between two SOAP spectra $\vec{a}$ and $\vec{b}$ can be calculated by:
\begin{equation}
d_{SOAP}\left(\vec{a},\vec{b}\right)=\sqrt{2-2\mathcal{K}\left(\vec{a},\vec{b}\right)}
\label{soap_linearkernel}
\end{equation}
where, with the SOAP power spectrum representation that we are using, $\mathcal{K}\left(\vec{a},\vec{b}\right)=\frac{\vec{a}\cdot\vec{b}}{\left\|\vec{a}\right\|\left\|\vec{b}\right\|}$.

For all systems, the SOAP descriptors were calculated under periodic boundary conditions along the \textit{xy} dimensions using nmax = 8 radial basis function and lmax = 8 maximum number of spherical harmonics.
The choice of the cut-off radius (rcut) determines the size and the shape of the neighborhood considered in characterizing the atomic environment for each SOAP center; in this work, we opted for rcut= $6.0$~\AA{}, the same cutoff adopted for the training of the Cu \textit{NN}-potential. The remaining parameters were set to default values of the DScribe library\cite{himanen2020dscribe} employed in this analysis.

\subsection{\textit{Bottom-up} SOAP analysis}

In the \textit{bottom-up} production analyses, we did not include the 2 bottom-most atomic layers (kept fixed) and the 3 closest neighbor layers of the bulk,  thus considering the first 1300 atoms of the Cu(211). This was done in order to guarantee a correct analysis of the atomic dynamics in the simulated metal surface.
We extracted the characteristic SOAP spectra for each of the 1300 Cu atoms in the surfaces from 250 snapshots taken from the last 75 ns of DPMD simulations (one every $\Delta t=300$ ps of simulation), corresponding in all simulated cases to the equilibrated-phase of the DPMD simulations.
In total, for the \textit{bottom-up} analysis of Figure~\ref{fig3}, we obtain a dataset of 325'000 SOAP power spectra in total.

The SOAP data output obtained from the DPMD trajectories consists of a set of feature vectors of high-dimensionality, that, while rich in information, are not convenient for visualization and classification.
To overcome these limitations, we opted to employ a widely used dimensionality reduction method, performing the Principal Component Analysis (PCA) algorithm in the implementation from scikit--learn~\cite{scikit-learn}, and to project the normalized SOAP vector on the principal components (PCs).
The data projected on the PCs is easier to visualize (in our case we used the first two PCs in Figure~\ref{fig3}c-d,f-g) and is convenient for finding patterns in the data.
More precisely, in the cases reported herein the first three PCs account for~$>98~\%$ of the total variance in the simulated systems.

The SOAP data extracted from the last 75~ns of the DPMD simulations (equilibrated-phase trajectories) have been then analyzed to identify the most visited atomic environments in these systems. In particular, unsuperfived clustering of the SOAP data has been performed using HDBSCAN*.\cite{HDBSCANCampello2013}
Density-based clustering algorithms identify clusters based on a search of high density peaks surrounded by regions where there is a lower density of points.
The minimum number of points in a neighborhood for a point to be recognized as a core point (min\_samples), did not impact the results. In particular, the results of Figure~\ref{fig3} are obtained using the default option, with automatic set to the same value as min\_cluster\_size and the default Excess of Mass (eom) as the cluster\_selection\_method.
As the clustering analysis is completely unsupervised, the detected clusters emerge \textit{bottom-up} directly from the second half (equilibrated-phase) of the DPMD trajectories.  

\subsection{Dictionary of SOAP atomic environments and \textit{top-down} analysis}

In Figures~\ref{fig4},\ref{fig5},\ref{fig6} we report a \textit{top-down} SOAP-based classification method that allows tracking and classifying all the atomic environments that appear in the systems during the equilibrium DPMD trajectories based on their similarity with references environments inserted in a SOAP data-set.
As a first step, we minimized the reference structures of Cu(111), (110), (210) and (211) surface slabs.
For all these minimized reference surfaces, we calculate the characteristic SOAP spectra of the atomic environments that compose them. In this case, we obtained a total of 22 characteristic SOAP spectra -- Figure~\ref{fig4}: 3 for (111), 5 for (110), 7 for (210) and 7 for (211). All these SOAP spectra have been then inserted into a unique SOAP dictionary of atomic environments, and used for our classification analysis. Noteworthy, this approach is flexible and general, and the dictionary can be in principle expanded by adding other SOAP spectra characteristic of other atomic environments if one wants to compare also other surface systems.
We then hierarchically classified all of the dictionary entries using the SOAP distance\cite{SOAPDistance} and by using the \textit{complete} method of the linkage algorithm, which conserves the greater distances between the newly formed cluster and the other elements.
In this analysis, our main aim is to compare the local atomic environments that emerge during the DPMD simulations and to classify these environments using the dictionary described in the dendrogram of Figure~\ref{fig4}c as a reference (on the right, in black: for this analysis we use a criterion of $d_{SOAP} \geq 0.01$).
For each atom in each simulation, we evaluated the SOAP distance ($d_{SOAP}$) from each dictionary entry. We then assigned each atom to the dictionary element with the lowest distance from it (\textit{i.e.}, smallest $d_{SOAP}$).
We applied this process to each atom (2304 for the Cu(210), 2400 for all other surfaces) along the entire 150 ns of DPMD simulations (501 frames, one every $\Delta t=300$ ps), thus obtaining 1'202'800 SOAP spectra for Cu(111), (211), (110) surfaces, and 1'154'304 SOAP spectra for Cu(210).
This allowed us to monitor the microscopic equilibration in all surface models (Figures~\ref{fig5},\ref{fig6}b).
For all cases, we took the second half of the simulations (the last 75~ns and 250 frames of DPMD) as representative of the equilibrium configuration of the surfaces.
The transition matrices of Figures~\ref{fig5},\ref{fig6}b are calculated by accumulating for each environment the number of changes to other clusters (or to self), and then by normalizing each row in the matrix to have it summing to 100, obtaining conditional probabilities for atoms belonging to a given class of environments to undergo transition into a new environments class (o to remain self) in $\Delta t=300$ ps.
Since in this analysis we are primarily interested in the mutability and transformations of the surface, only the SOAP data for the first topmost surface (s) layers of atoms are retained, while all atoms belonging to bulk or to the ss$^{h}$ environments at DPMD simulation start (in the ideal surfaces) are not considered in the analysis.

\section{Supplementary Material}
Supplementary Material contains details about: the validation of the \textit{NN}-potential; time-series of the populations of the clusters determined by the \textit{bottom-up} SOAP analysis for all the surfaces investigated; a coordination and \textit{top-down} SOAP based analyses of the reconstructed (110)(1$\times$2), (110)(1$\times$3) and (211)(2$\times$1) surfaces; the radial distribution function of the (111),(110),(211) and (210) surfaces at T=0, 500 and 700 K; additional details on the \textit{bootm-up} and \textit{top-down} SOAP based analyses.    

\section{Acknowledgements}
G.M.P. acknowledges the funding received by the ERC under the European Union’s Horizon 2020 research and innovation program (grant agreement no. 818776 - DYNAPOL) and by the Swiss National Science Foundation (SNSF grants IZLIZ2\_183336).The authors also acknowledge the computational resources provided by the Swiss National Supercomputing Center (CSCS), by CINECA and by HPC@POLITO (http://www.hpc.polito.it).

\section{Data availability}
Details on the procedures for the training of the neural network potential and on the simulations’ setup, along with additional simulation data, are provided in the Methods section and in the Supplementary Information file. Complete data and materials pertaining to the molecular simulations and data analyses conducted herein (input files, model files, raw data, analysis tools, etc.)
are available at https://github.com/GMPavanLab/DynMetSurf (this link will be replaced with a definitive Zenodo link upon acceptance of the final version of this paper). Other information needed is available from the corresponding author upon reasonable request.

\section{Author contributions}
M.C., D.P. and M.D.P. developed the molecular models. M.C. performed the simulations. M.C., D.P., D.R, L.P, M.D.P. and G.M.P. analysed the results. G.M.P. conceived this research and supervised the work. M.C., D.P., D.R., M.D.P and G.M.P wrote the manuscript.

\section{Competing interests statement}
The authors declare no competing interests.


%
%

%


\bibliography{bibliography}

\begin{thebibliography}{104}%
\makeatletter
\providecommand \@ifxundefined [1]{%
 \@ifx{#1\undefined}
}%
\providecommand \@ifnum [1]{%
 \ifnum #1\expandafter \@firstoftwo
 \else \expandafter \@secondoftwo
 \fi
}%
\providecommand \@ifx [1]{%
 \ifx #1\expandafter \@firstoftwo
 \else \expandafter \@secondoftwo
 \fi
}%
\providecommand \natexlab [1]{#1}%
\providecommand \enquote  [1]{``#1''}%
\providecommand \bibnamefont  [1]{#1}%
\providecommand \bibfnamefont [1]{#1}%
\providecommand \citenamefont [1]{#1}%
\providecommand \href@noop [0]{\@secondoftwo}%
\providecommand \href [0]{\begingroup \@sanitize@url \@href}%
\providecommand \@href[1]{\@@startlink{#1}\@@href}%
\providecommand \@@href[1]{\endgroup#1\@@endlink}%
\providecommand \@sanitize@url [0]{\catcode `\\12\catcode `\$12\catcode
  `\&12\catcode `\#12\catcode `\^12\catcode `\_12\catcode `\%12\relax}%
\providecommand \@@startlink[1]{}%
\providecommand \@@endlink[0]{}%
\providecommand \url  [0]{\begingroup\@sanitize@url \@url }%
\providecommand \@url [1]{\endgroup\@href {#1}{\urlprefix }}%
\providecommand \urlprefix  [0]{URL }%
\providecommand \Eprint [0]{\href }%
\providecommand \doibase [0]{http://dx.doi.org/}%
\providecommand \selectlanguage [0]{\@gobble}%
\providecommand \bibinfo  [0]{\@secondoftwo}%
\providecommand \bibfield  [0]{\@secondoftwo}%
\providecommand \translation [1]{[#1]}%
\providecommand \BibitemOpen [0]{}%
\providecommand \bibitemStop [0]{}%
\providecommand \bibitemNoStop [0]{.\EOS\space}%
\providecommand \EOS [0]{\spacefactor3000\relax}%
\providecommand \BibitemShut  [1]{\csname bibitem#1\endcsname}%
\let\auto@bib@innerbib\@empty
\bibitem [{\citenamefont {Desjonqueres}\ and\ \citenamefont
  {Spanjaard}(1996)}]{desjonqueres1996concepts}%
  \BibitemOpen
  \bibfield  {author} {\bibinfo {author} {\bibfnamefont {M.-C.}\ \bibnamefont
  {Desjonqueres}}\ and\ \bibinfo {author} {\bibfnamefont {D.}~\bibnamefont
  {Spanjaard}},\ }\href@noop {} {\emph {\bibinfo {title} {Concepts in Surface
  Physics: 2eme {\'e}dition}}},\ Vol.~\bibinfo {volume} {30}\ (\bibinfo
  {publisher} {Springer Science \& Business Media},\ \bibinfo {year}
  {1996})\BibitemShut {NoStop}%
\bibitem [{\citenamefont {Spencer}(1986)}]{spencer1986stable}%
  \BibitemOpen
  \bibfield  {author} {\bibinfo {author} {\bibfnamefont {M.}~\bibnamefont
  {Spencer}},\ }\href@noop {} {\bibfield  {journal} {\bibinfo  {journal}
  {Nature}\ }\textbf {\bibinfo {volume} {323}},\ \bibinfo {pages} {685}
  (\bibinfo {year} {1986})}\BibitemShut {NoStop}%
\bibitem [{\citenamefont {Jayanthi}, \citenamefont {Tosatti},\ and\
  \citenamefont {Pietronero}(1985)}]{jayanthi1985surface}%
  \BibitemOpen
  \bibfield  {author} {\bibinfo {author} {\bibfnamefont {C.}~\bibnamefont
  {Jayanthi}}, \bibinfo {author} {\bibfnamefont {E.}~\bibnamefont {Tosatti}}, \
  and\ \bibinfo {author} {\bibfnamefont {L.}~\bibnamefont {Pietronero}},\
  }\href@noop {} {\bibfield  {journal} {\bibinfo  {journal} {Phys. Rev. B}\
  }\textbf {\bibinfo {volume} {31}},\ \bibinfo {pages} {3456} (\bibinfo {year}
  {1985})}\BibitemShut {NoStop}%
\bibitem [{\citenamefont {Bernasconi}\ and\ \citenamefont
  {Tosatti}(1993)}]{bernasconi1993reconstruction}%
  \BibitemOpen
  \bibfield  {author} {\bibinfo {author} {\bibfnamefont {M.}~\bibnamefont
  {Bernasconi}}\ and\ \bibinfo {author} {\bibfnamefont {E.}~\bibnamefont
  {Tosatti}},\ }\href@noop {} {\bibfield  {journal} {\bibinfo  {journal} {Surf.
  Sci. Rep.}\ }\textbf {\bibinfo {volume} {17}},\ \bibinfo {pages} {363}
  (\bibinfo {year} {1993})}\BibitemShut {NoStop}%
\bibitem [{\citenamefont {Trayanov}, \citenamefont {Levi},\ and\ \citenamefont
  {Tosatti}(1990)}]{trayanov1990anisotropic}%
  \BibitemOpen
  \bibfield  {author} {\bibinfo {author} {\bibfnamefont {A.}~\bibnamefont
  {Trayanov}}, \bibinfo {author} {\bibfnamefont {A.}~\bibnamefont {Levi}}, \
  and\ \bibinfo {author} {\bibfnamefont {E.}~\bibnamefont {Tosatti}},\
  }\href@noop {} {\bibfield  {journal} {\bibinfo  {journal} {Surface Science}\
  }\textbf {\bibinfo {volume} {233}},\ \bibinfo {pages} {184} (\bibinfo {year}
  {1990})}\BibitemShut {NoStop}%
\bibitem [{\citenamefont {D{\"u}rr}, \citenamefont {Schneider},\ and\
  \citenamefont {Fauster}(1991)}]{durr1991anomalous}%
  \BibitemOpen
  \bibfield  {author} {\bibinfo {author} {\bibfnamefont {H.}~\bibnamefont
  {D{\"u}rr}}, \bibinfo {author} {\bibfnamefont {R.}~\bibnamefont {Schneider}},
  \ and\ \bibinfo {author} {\bibfnamefont {T.}~\bibnamefont {Fauster}},\
  }\href@noop {} {\bibfield  {journal} {\bibinfo  {journal} {Physical Review
  B}\ }\textbf {\bibinfo {volume} {43}},\ \bibinfo {pages} {12187} (\bibinfo
  {year} {1991})}\BibitemShut {NoStop}%
\bibitem [{\citenamefont {Merikoski}\ \emph {et~al.}(1994)\citenamefont
  {Merikoski}, \citenamefont {H{\"a}kkinen}, \citenamefont {Manninen},
  \citenamefont {Timonen},\ and\ \citenamefont
  {Kaski}}]{merikoski1994disordering}%
  \BibitemOpen
  \bibfield  {author} {\bibinfo {author} {\bibfnamefont {J.}~\bibnamefont
  {Merikoski}}, \bibinfo {author} {\bibfnamefont {H.}~\bibnamefont
  {H{\"a}kkinen}}, \bibinfo {author} {\bibfnamefont {M.}~\bibnamefont
  {Manninen}}, \bibinfo {author} {\bibfnamefont {J.}~\bibnamefont {Timonen}}, \
  and\ \bibinfo {author} {\bibfnamefont {K.}~\bibnamefont {Kaski}},\
  }\href@noop {} {\bibfield  {journal} {\bibinfo  {journal} {International
  Journal of Modern Physics B}\ }\textbf {\bibinfo {volume} {8}},\ \bibinfo
  {pages} {3175} (\bibinfo {year} {1994})}\BibitemShut {NoStop}%
\bibitem [{\citenamefont {Zeybek}(2006)}]{zeybek2006thermal}%
  \BibitemOpen
  \bibfield  {author} {\bibinfo {author} {\bibfnamefont {O.}~\bibnamefont
  {Zeybek}},\ }\href@noop {} {\bibfield  {journal} {\bibinfo  {journal} {Solid
  state communications}\ }\textbf {\bibinfo {volume} {139}},\ \bibinfo {pages}
  {339} (\bibinfo {year} {2006})}\BibitemShut {NoStop}%
\bibitem [{\citenamefont {Xie}\ \emph {et~al.}(2013)\citenamefont {Xie},
  \citenamefont {Choi}, \citenamefont {Xia},\ and\ \citenamefont
  {Xia}}]{xie2013catalysis}%
  \BibitemOpen
  \bibfield  {author} {\bibinfo {author} {\bibfnamefont {S.}~\bibnamefont
  {Xie}}, \bibinfo {author} {\bibfnamefont {S.-I.}\ \bibnamefont {Choi}},
  \bibinfo {author} {\bibfnamefont {X.}~\bibnamefont {Xia}}, \ and\ \bibinfo
  {author} {\bibfnamefont {Y.}~\bibnamefont {Xia}},\ }\href@noop {} {\bibfield
  {journal} {\bibinfo  {journal} {Curr. Opin. Chem. Eng.}\ }\textbf {\bibinfo
  {volume} {2}},\ \bibinfo {pages} {142} (\bibinfo {year} {2013})}\BibitemShut
  {NoStop}%
\bibitem [{\citenamefont {Dattila}, \citenamefont {Garc{\i}\'{a}-Muelas},\ and\
  \citenamefont {L\'{o}pez}(2020)}]{dattila2020active}%
  \BibitemOpen
  \bibfield  {author} {\bibinfo {author} {\bibfnamefont {F.}~\bibnamefont
  {Dattila}}, \bibinfo {author} {\bibfnamefont {R.}~\bibnamefont
  {Garc{\i}\'{a}-Muelas}}, \ and\ \bibinfo {author} {\bibfnamefont
  {N.}~\bibnamefont {L\'{o}pez}},\ }\href@noop {} {\bibfield  {journal}
  {\bibinfo  {journal} {ACS Energy Lett.}\ }\textbf {\bibinfo {volume} {5}},\
  \bibinfo {pages} {3176} (\bibinfo {year} {2020})}\BibitemShut {NoStop}%
\bibitem [{\citenamefont {Yamakov}\ \emph {et~al.}(2004)\citenamefont
  {Yamakov}, \citenamefont {Wolf}, \citenamefont {Phillpot}, \citenamefont
  {Mukherjee},\ and\ \citenamefont {Gleiter}}]{yamakov2004deformation}%
  \BibitemOpen
  \bibfield  {author} {\bibinfo {author} {\bibfnamefont {V.}~\bibnamefont
  {Yamakov}}, \bibinfo {author} {\bibfnamefont {D.}~\bibnamefont {Wolf}},
  \bibinfo {author} {\bibfnamefont {S.}~\bibnamefont {Phillpot}}, \bibinfo
  {author} {\bibfnamefont {A.}~\bibnamefont {Mukherjee}}, \ and\ \bibinfo
  {author} {\bibfnamefont {H.}~\bibnamefont {Gleiter}},\ }\href@noop {}
  {\bibfield  {journal} {\bibinfo  {journal} {Nat. Mat.}\ }\textbf {\bibinfo
  {volume} {3}},\ \bibinfo {pages} {43} (\bibinfo {year} {2004})}\BibitemShut
  {NoStop}%
\bibitem [{\citenamefont {Zepeda-Ruiz}\ \emph {et~al.}(2017)\citenamefont
  {Zepeda-Ruiz}, \citenamefont {Stukowski}, \citenamefont {Oppelstrup},\ and\
  \citenamefont {Bulatov}}]{zepeda2017probing}%
  \BibitemOpen
  \bibfield  {author} {\bibinfo {author} {\bibfnamefont {L.~A.}\ \bibnamefont
  {Zepeda-Ruiz}}, \bibinfo {author} {\bibfnamefont {A.}~\bibnamefont
  {Stukowski}}, \bibinfo {author} {\bibfnamefont {T.}~\bibnamefont
  {Oppelstrup}}, \ and\ \bibinfo {author} {\bibfnamefont {V.~V.}\ \bibnamefont
  {Bulatov}},\ }\href@noop {} {\bibfield  {journal} {\bibinfo  {journal}
  {Nature}\ }\textbf {\bibinfo {volume} {550}},\ \bibinfo {pages} {492}
  (\bibinfo {year} {2017})}\BibitemShut {NoStop}%
\bibitem [{\citenamefont {Wang}\ \emph {et~al.}(2021)\citenamefont {Wang},
  \citenamefont {Zheng}, \citenamefont {Shinzato}, \citenamefont {Fang},
  \citenamefont {He}, \citenamefont {Zhong}, \citenamefont {Wang},
  \citenamefont {Ogata},\ and\ \citenamefont {Mao}}]{Wang2021}%
  \BibitemOpen
  \bibfield  {author} {\bibinfo {author} {\bibfnamefont {X.}~\bibnamefont
  {Wang}}, \bibinfo {author} {\bibfnamefont {S.}~\bibnamefont {Zheng}},
  \bibinfo {author} {\bibfnamefont {S.}~\bibnamefont {Shinzato}}, \bibinfo
  {author} {\bibfnamefont {Z.}~\bibnamefont {Fang}}, \bibinfo {author}
  {\bibfnamefont {Y.}~\bibnamefont {He}}, \bibinfo {author} {\bibfnamefont
  {L.}~\bibnamefont {Zhong}}, \bibinfo {author} {\bibfnamefont
  {C.}~\bibnamefont {Wang}}, \bibinfo {author} {\bibfnamefont {S.}~\bibnamefont
  {Ogata}}, \ and\ \bibinfo {author} {\bibfnamefont {S.~X.}\ \bibnamefont
  {Mao}},\ }\href {\doibase 10.1038/s41467-021-25542-2} {\bibfield  {journal}
  {\bibinfo  {journal} {Nat. Commun.}\ }\textbf {\bibinfo {volume} {12}},\
  \bibinfo {pages} {5237} (\bibinfo {year} {2021})}\BibitemShut {NoStop}%
\bibitem [{\citenamefont {N{\o}rskov}\ \emph {et~al.}(2009)\citenamefont
  {N{\o}rskov}, \citenamefont {Bligaard}, \citenamefont {Rossmeisl},\ and\
  \citenamefont {Christensen}}]{Norskov2009}%
  \BibitemOpen
  \bibfield  {author} {\bibinfo {author} {\bibfnamefont {J.~K.}\ \bibnamefont
  {N{\o}rskov}}, \bibinfo {author} {\bibfnamefont {T.}~\bibnamefont
  {Bligaard}}, \bibinfo {author} {\bibfnamefont {J.}~\bibnamefont {Rossmeisl}},
  \ and\ \bibinfo {author} {\bibfnamefont {C.~H.}\ \bibnamefont
  {Christensen}},\ }\href {\doibase 10.1038/nchem.121} {\bibfield  {journal}
  {\bibinfo  {journal} {Nat. Chem.}\ }\textbf {\bibinfo {volume} {1}},\
  \bibinfo {pages} {37} (\bibinfo {year} {2009})}\BibitemShut {NoStop}%
\bibitem [{\citenamefont {Nørskov}\ \emph {et~al.}(2011)\citenamefont
  {Nørskov}, \citenamefont {Abild-Pedersen}, \citenamefont {Studt},\ and\
  \citenamefont {Bligaard}}]{Norskov2011}%
  \BibitemOpen
  \bibfield  {author} {\bibinfo {author} {\bibfnamefont {J.~K.}\ \bibnamefont
  {Nørskov}}, \bibinfo {author} {\bibfnamefont {F.}~\bibnamefont
  {Abild-Pedersen}}, \bibinfo {author} {\bibfnamefont {F.}~\bibnamefont
  {Studt}}, \ and\ \bibinfo {author} {\bibfnamefont {T.}~\bibnamefont
  {Bligaard}},\ }\href {\doibase 10.1073/pnas.1006652108} {\bibfield  {journal}
  {\bibinfo  {journal} {Proc. Natl. Acad. Sci.}\ }\textbf {\bibinfo {volume}
  {108}},\ \bibinfo {pages} {937} (\bibinfo {year} {2011})}\BibitemShut
  {NoStop}%
\bibitem [{\citenamefont {Calle-Vallejo}\ \emph {et~al.}(2015)\citenamefont
  {Calle-Vallejo}, \citenamefont {Tymoczko}, \citenamefont {Colic},
  \citenamefont {Vu}, \citenamefont {Pohl}, \citenamefont {Morgenstern},
  \citenamefont {Loffreda}, \citenamefont {Sautet}, \citenamefont {Schuhmann},\
  and\ \citenamefont {Bandarenka}}]{CalleVallejo2015}%
  \BibitemOpen
  \bibfield  {author} {\bibinfo {author} {\bibfnamefont {F.}~\bibnamefont
  {Calle-Vallejo}}, \bibinfo {author} {\bibfnamefont {J.}~\bibnamefont
  {Tymoczko}}, \bibinfo {author} {\bibfnamefont {V.}~\bibnamefont {Colic}},
  \bibinfo {author} {\bibfnamefont {Q.~H.}\ \bibnamefont {Vu}}, \bibinfo
  {author} {\bibfnamefont {M.~D.}\ \bibnamefont {Pohl}}, \bibinfo {author}
  {\bibfnamefont {K.}~\bibnamefont {Morgenstern}}, \bibinfo {author}
  {\bibfnamefont {D.}~\bibnamefont {Loffreda}}, \bibinfo {author}
  {\bibfnamefont {P.}~\bibnamefont {Sautet}}, \bibinfo {author} {\bibfnamefont
  {W.}~\bibnamefont {Schuhmann}}, \ and\ \bibinfo {author} {\bibfnamefont
  {A.~S.}\ \bibnamefont {Bandarenka}},\ }\href {\doibase
  10.1126/science.aab3501} {\bibfield  {journal} {\bibinfo  {journal}
  {Science}\ }\textbf {\bibinfo {volume} {350}},\ \bibinfo {pages} {185}
  (\bibinfo {year} {2015})}\BibitemShut {NoStop}%
\bibitem [{\citenamefont {Zhong}\ \emph {et~al.}(2020)\citenamefont {Zhong},
  \citenamefont {Tran}, \citenamefont {Min}, \citenamefont {Wang},
  \citenamefont {Wang}, \citenamefont {Dinh}, \citenamefont {De~Luna},
  \citenamefont {Yu}, \citenamefont {Rasouli}, \citenamefont {Brodersen},
  \citenamefont {Sun}, \citenamefont {Voznyy}, \citenamefont {Tan},
  \citenamefont {Askerka}, \citenamefont {Che}, \citenamefont {Liu},
  \citenamefont {Seifitokaldani}, \citenamefont {Pang}, \citenamefont {Lo},
  \citenamefont {Ip}, \citenamefont {Ulissi},\ and\ \citenamefont
  {Sargent}}]{Zhong2020}%
  \BibitemOpen
  \bibfield  {author} {\bibinfo {author} {\bibfnamefont {M.}~\bibnamefont
  {Zhong}}, \bibinfo {author} {\bibfnamefont {K.}~\bibnamefont {Tran}},
  \bibinfo {author} {\bibfnamefont {Y.}~\bibnamefont {Min}}, \bibinfo {author}
  {\bibfnamefont {C.}~\bibnamefont {Wang}}, \bibinfo {author} {\bibfnamefont
  {Z.}~\bibnamefont {Wang}}, \bibinfo {author} {\bibfnamefont {C.-T.}\
  \bibnamefont {Dinh}}, \bibinfo {author} {\bibfnamefont {P.}~\bibnamefont
  {De~Luna}}, \bibinfo {author} {\bibfnamefont {Z.}~\bibnamefont {Yu}},
  \bibinfo {author} {\bibfnamefont {A.~S.}\ \bibnamefont {Rasouli}}, \bibinfo
  {author} {\bibfnamefont {P.}~\bibnamefont {Brodersen}}, \bibinfo {author}
  {\bibfnamefont {S.}~\bibnamefont {Sun}}, \bibinfo {author} {\bibfnamefont
  {O.}~\bibnamefont {Voznyy}}, \bibinfo {author} {\bibfnamefont {C.-S.}\
  \bibnamefont {Tan}}, \bibinfo {author} {\bibfnamefont {M.}~\bibnamefont
  {Askerka}}, \bibinfo {author} {\bibfnamefont {F.}~\bibnamefont {Che}},
  \bibinfo {author} {\bibfnamefont {M.}~\bibnamefont {Liu}}, \bibinfo {author}
  {\bibfnamefont {A.}~\bibnamefont {Seifitokaldani}}, \bibinfo {author}
  {\bibfnamefont {Y.}~\bibnamefont {Pang}}, \bibinfo {author} {\bibfnamefont
  {S.-C.}\ \bibnamefont {Lo}}, \bibinfo {author} {\bibfnamefont
  {A.}~\bibnamefont {Ip}}, \bibinfo {author} {\bibfnamefont {Z.}~\bibnamefont
  {Ulissi}}, \ and\ \bibinfo {author} {\bibfnamefont {E.~H.}\ \bibnamefont
  {Sargent}},\ }\href {\doibase 10.1038/s41586-020-2242-8} {\bibfield
  {journal} {\bibinfo  {journal} {Nature}\ }\textbf {\bibinfo {volume} {581}},\
  \bibinfo {pages} {178} (\bibinfo {year} {2020})}\BibitemShut {NoStop}%
\bibitem [{\citenamefont {Chen}, \citenamefont {Xu},\ and\ \citenamefont
  {Mavrikakis}(2021)}]{Mavrikakis2021}%
  \BibitemOpen
  \bibfield  {author} {\bibinfo {author} {\bibfnamefont {B.~W.~J.}\
  \bibnamefont {Chen}}, \bibinfo {author} {\bibfnamefont {L.}~\bibnamefont
  {Xu}}, \ and\ \bibinfo {author} {\bibfnamefont {M.}~\bibnamefont
  {Mavrikakis}},\ }\href {\doibase 10.1021/acs.chemrev.0c01060} {\bibfield
  {journal} {\bibinfo  {journal} {Chem. Rev.}\ }\textbf {\bibinfo {volume}
  {121}},\ \bibinfo {pages} {1007} (\bibinfo {year} {2021})}\BibitemShut
  {NoStop}%
\bibitem [{\citenamefont {Gazzarrini}, \citenamefont {Rossi},\ and\
  \citenamefont {Baletto}(2021)}]{gazzarrini2021born}%
  \BibitemOpen
  \bibfield  {author} {\bibinfo {author} {\bibfnamefont {E.}~\bibnamefont
  {Gazzarrini}}, \bibinfo {author} {\bibfnamefont {K.}~\bibnamefont {Rossi}}, \
  and\ \bibinfo {author} {\bibfnamefont {F.}~\bibnamefont {Baletto}},\ }\href
  {\doibase 10.1039/D0NR07889A} {\bibfield  {journal} {\bibinfo  {journal}
  {Nanoscale}\ }\textbf {\bibinfo {volume} {13}},\ \bibinfo {pages} {5857}
  (\bibinfo {year} {2021})}\BibitemShut {NoStop}%
\bibitem [{\citenamefont {Nelli}, \citenamefont {Pietrucci},\ and\
  \citenamefont {Ferrando}(2021)}]{Nelli2021}%
  \BibitemOpen
  \bibfield  {author} {\bibinfo {author} {\bibfnamefont {D.}~\bibnamefont
  {Nelli}}, \bibinfo {author} {\bibfnamefont {F.}~\bibnamefont {Pietrucci}}, \
  and\ \bibinfo {author} {\bibfnamefont {R.}~\bibnamefont {Ferrando}},\ }\href
  {\doibase 10.1063/5.0060236} {\bibfield  {journal} {\bibinfo  {journal} {J.
  Chem. Phys.}\ }\textbf {\bibinfo {volume} {155}},\ \bibinfo {pages} {144304}
  (\bibinfo {year} {2021})}\BibitemShut {NoStop}%
\bibitem [{\citenamefont {Crippa}\ \emph {et~al.}(2022)\citenamefont {Crippa},
  \citenamefont {Perego}, \citenamefont {de~Marco},\ and\ \citenamefont
  {Pavan}}]{crippa2022molecular}%
  \BibitemOpen
  \bibfield  {author} {\bibinfo {author} {\bibfnamefont {M.}~\bibnamefont
  {Crippa}}, \bibinfo {author} {\bibfnamefont {C.}~\bibnamefont {Perego}},
  \bibinfo {author} {\bibfnamefont {A.~L.}\ \bibnamefont {de~Marco}}, \ and\
  \bibinfo {author} {\bibfnamefont {G.~M.}\ \bibnamefont {Pavan}},\ }\href@noop
  {} {\bibfield  {journal} {\bibinfo  {journal} {Nat. Commun.}\ }\textbf
  {\bibinfo {volume} {13}},\ \bibinfo {pages} {1} (\bibinfo {year}
  {2022})}\BibitemShut {NoStop}%
\bibitem [{\citenamefont {Gasparotto}\ \emph {et~al.}(2019)\citenamefont
  {Gasparotto}, \citenamefont {Bochicchio}, \citenamefont {Ceriotti},\ and\
  \citenamefont {Pavan}}]{gasparotto2019identifying}%
  \BibitemOpen
  \bibfield  {author} {\bibinfo {author} {\bibfnamefont {P.}~\bibnamefont
  {Gasparotto}}, \bibinfo {author} {\bibfnamefont {D.}~\bibnamefont
  {Bochicchio}}, \bibinfo {author} {\bibfnamefont {M.}~\bibnamefont
  {Ceriotti}}, \ and\ \bibinfo {author} {\bibfnamefont {G.~M.}\ \bibnamefont
  {Pavan}},\ }\href@noop {} {\bibfield  {journal} {\bibinfo  {journal} {J.
  Phys. Chem. B}\ }\textbf {\bibinfo {volume} {124}},\ \bibinfo {pages} {589}
  (\bibinfo {year} {2019})}\BibitemShut {NoStop}%
\bibitem [{\citenamefont {Gardin}\ \emph {et~al.}(2022)\citenamefont {Gardin},
  \citenamefont {Perego}, \citenamefont {Doni},\ and\ \citenamefont
  {Pavan}}]{gardin2022classifying}%
  \BibitemOpen
  \bibfield  {author} {\bibinfo {author} {\bibfnamefont {A.}~\bibnamefont
  {Gardin}}, \bibinfo {author} {\bibfnamefont {C.}~\bibnamefont {Perego}},
  \bibinfo {author} {\bibfnamefont {G.}~\bibnamefont {Doni}}, \ and\ \bibinfo
  {author} {\bibfnamefont {G.~M.}\ \bibnamefont {Pavan}},\ }\href@noop {}
  {\bibfield  {journal} {\bibinfo  {journal} {Commun. Chem.}\ }\textbf
  {\bibinfo {volume} {5}},\ \bibinfo {pages} {1} (\bibinfo {year}
  {2022})}\BibitemShut {NoStop}%
\bibitem [{\citenamefont {Bochicchio}, \citenamefont {Salvalaglio},\ and\
  \citenamefont {Pavan}(2017)}]{bochicchio2017into}%
  \BibitemOpen
  \bibfield  {author} {\bibinfo {author} {\bibfnamefont {D.}~\bibnamefont
  {Bochicchio}}, \bibinfo {author} {\bibfnamefont {M.}~\bibnamefont
  {Salvalaglio}}, \ and\ \bibinfo {author} {\bibfnamefont {G.~M.}\ \bibnamefont
  {Pavan}},\ }\href@noop {} {\bibfield  {journal} {\bibinfo  {journal} {Nat.
  Commun.}\ }\textbf {\bibinfo {volume} {8}},\ \bibinfo {pages} {1} (\bibinfo
  {year} {2017})}\BibitemShut {NoStop}%
\bibitem [{\citenamefont {Behler}\ and\ \citenamefont
  {Parrinello}(2007)}]{BehlerParrinello2007}%
  \BibitemOpen
  \bibfield  {author} {\bibinfo {author} {\bibfnamefont {J.}~\bibnamefont
  {Behler}}\ and\ \bibinfo {author} {\bibfnamefont {M.}~\bibnamefont
  {Parrinello}},\ }\href {\doibase 10.1103/PhysRevLett.98.146401} {\bibfield
  {journal} {\bibinfo  {journal} {Phys. Rev. Lett.}\ }\textbf {\bibinfo
  {volume} {98}},\ \bibinfo {pages} {146401} (\bibinfo {year}
  {2007})}\BibitemShut {NoStop}%
\bibitem [{\citenamefont {Bart\'ok}\ \emph {et~al.}(2010)\citenamefont
  {Bart\'ok}, \citenamefont {Payne}, \citenamefont {Kondor},\ and\
  \citenamefont {Cs\'anyi}}]{Csanyi2010}%
  \BibitemOpen
  \bibfield  {author} {\bibinfo {author} {\bibfnamefont {A.~P.}\ \bibnamefont
  {Bart\'ok}}, \bibinfo {author} {\bibfnamefont {M.~C.}\ \bibnamefont {Payne}},
  \bibinfo {author} {\bibfnamefont {R.}~\bibnamefont {Kondor}}, \ and\ \bibinfo
  {author} {\bibfnamefont {G.}~\bibnamefont {Cs\'anyi}},\ }\href {\doibase
  10.1103/PhysRevLett.104.136403} {\bibfield  {journal} {\bibinfo  {journal}
  {Phys. Rev. Lett.}\ }\textbf {\bibinfo {volume} {104}},\ \bibinfo {pages}
  {136403} (\bibinfo {year} {2010})}\BibitemShut {NoStop}%
\bibitem [{\citenamefont {Doerr}\ \emph {et~al.}(2021)\citenamefont {Doerr},
  \citenamefont {Majewski}, \citenamefont {Pérez}, \citenamefont {Krämer},
  \citenamefont {Clementi}, \citenamefont {Noe}, \citenamefont {Giorgino},\
  and\ \citenamefont {De~Fabritiis}}]{DeFabritiis2021TorchMD}%
  \BibitemOpen
  \bibfield  {author} {\bibinfo {author} {\bibfnamefont {S.}~\bibnamefont
  {Doerr}}, \bibinfo {author} {\bibfnamefont {M.}~\bibnamefont {Majewski}},
  \bibinfo {author} {\bibfnamefont {A.}~\bibnamefont {Pérez}}, \bibinfo
  {author} {\bibfnamefont {A.}~\bibnamefont {Krämer}}, \bibinfo {author}
  {\bibfnamefont {C.}~\bibnamefont {Clementi}}, \bibinfo {author}
  {\bibfnamefont {F.}~\bibnamefont {Noe}}, \bibinfo {author} {\bibfnamefont
  {T.}~\bibnamefont {Giorgino}}, \ and\ \bibinfo {author} {\bibfnamefont
  {G.}~\bibnamefont {De~Fabritiis}},\ }\href {\doibase
  10.1021/acs.jctc.0c01343} {\bibfield  {journal} {\bibinfo  {journal} {J.
  Chem. Theory Comput.}\ }\textbf {\bibinfo {volume} {17}},\ \bibinfo {pages}
  {2355} (\bibinfo {year} {2021})}\BibitemShut {NoStop}%
\bibitem [{\citenamefont {Schütt}\ \emph {et~al.}(2018)\citenamefont
  {Schütt}, \citenamefont {Sauceda}, \citenamefont {Kindermans}, \citenamefont
  {Tkatchenko},\ and\ \citenamefont {Müller}}]{SchNet2018}%
  \BibitemOpen
  \bibfield  {author} {\bibinfo {author} {\bibfnamefont {K.~T.}\ \bibnamefont
  {Schütt}}, \bibinfo {author} {\bibfnamefont {H.~E.}\ \bibnamefont
  {Sauceda}}, \bibinfo {author} {\bibfnamefont {P.-J.}\ \bibnamefont
  {Kindermans}}, \bibinfo {author} {\bibfnamefont {A.}~\bibnamefont
  {Tkatchenko}}, \ and\ \bibinfo {author} {\bibfnamefont {K.-R.}\ \bibnamefont
  {Müller}},\ }\href {\doibase 10.1063/1.5019779} {\bibfield  {journal}
  {\bibinfo  {journal} {J. Chem. Phys.}\ }\textbf {\bibinfo {volume} {148}},\
  \bibinfo {pages} {241722} (\bibinfo {year} {2018})}\BibitemShut {NoStop}%
\bibitem [{\citenamefont {Unke}\ and\ \citenamefont
  {Meuwly}(2019)}]{PhysNet2019}%
  \BibitemOpen
  \bibfield  {author} {\bibinfo {author} {\bibfnamefont {O.~T.}\ \bibnamefont
  {Unke}}\ and\ \bibinfo {author} {\bibfnamefont {M.}~\bibnamefont {Meuwly}},\
  }\href {\doibase 10.1021/acs.jctc.9b00181} {\bibfield  {journal} {\bibinfo
  {journal} {J. Chem. Theory Comput.}\ }\textbf {\bibinfo {volume} {15}},\
  \bibinfo {pages} {3678} (\bibinfo {year} {2019})}\BibitemShut {NoStop}%
\bibitem [{\citenamefont {Li}, \citenamefont {Kermode},\ and\ \citenamefont
  {De~Vita}(2015)}]{DeVita2015}%
  \BibitemOpen
  \bibfield  {author} {\bibinfo {author} {\bibfnamefont {Z.}~\bibnamefont
  {Li}}, \bibinfo {author} {\bibfnamefont {J.~R.}\ \bibnamefont {Kermode}}, \
  and\ \bibinfo {author} {\bibfnamefont {A.}~\bibnamefont {De~Vita}},\ }\href
  {\doibase 10.1103/PhysRevLett.114.096405} {\bibfield  {journal} {\bibinfo
  {journal} {Phys. Rev. Lett.}\ }\textbf {\bibinfo {volume} {114}},\ \bibinfo
  {pages} {096405} (\bibinfo {year} {2015})}\BibitemShut {NoStop}%
\bibitem [{\citenamefont {Wang}\ \emph {et~al.}(2018)\citenamefont {Wang},
  \citenamefont {Zhang}, \citenamefont {Han},\ and\ \citenamefont
  {E}}]{deepMD}%
  \BibitemOpen
  \bibfield  {author} {\bibinfo {author} {\bibfnamefont {H.}~\bibnamefont
  {Wang}}, \bibinfo {author} {\bibfnamefont {L.}~\bibnamefont {Zhang}},
  \bibinfo {author} {\bibfnamefont {J.}~\bibnamefont {Han}}, \ and\ \bibinfo
  {author} {\bibfnamefont {W.}~\bibnamefont {E}},\ }\href {\doibase
  https://doi.org/10.1016/j.cpc.2018.03.016} {\bibfield  {journal} {\bibinfo
  {journal} {Comput. Phys. Commun.}\ }\textbf {\bibinfo {volume} {228}},\
  \bibinfo {pages} {178} (\bibinfo {year} {2018})}\BibitemShut {NoStop}%
\bibitem [{\citenamefont {Pun}\ \emph {et~al.}(2019)\citenamefont {Pun},
  \citenamefont {Batra}, \citenamefont {Ramprasad},\ and\ \citenamefont
  {Mishin}}]{Pun2019-PINN}%
  \BibitemOpen
  \bibfield  {author} {\bibinfo {author} {\bibfnamefont {G.~P.~P.}\
  \bibnamefont {Pun}}, \bibinfo {author} {\bibfnamefont {R.}~\bibnamefont
  {Batra}}, \bibinfo {author} {\bibfnamefont {R.}~\bibnamefont {Ramprasad}}, \
  and\ \bibinfo {author} {\bibfnamefont {Y.}~\bibnamefont {Mishin}},\ }\href
  {\doibase 10.1038/s41467-019-10343-5} {\bibfield  {journal} {\bibinfo
  {journal} {Nat. Commun.}\ }\textbf {\bibinfo {volume} {10}},\ \bibinfo
  {pages} {2339} (\bibinfo {year} {2019})}\BibitemShut {NoStop}%
\bibitem [{\citenamefont {Zuo}\ \emph {et~al.}(2020)\citenamefont {Zuo},
  \citenamefont {Chen}, \citenamefont {Li}, \citenamefont {Deng}, \citenamefont
  {Chen}, \citenamefont {Behler}, \citenamefont {Csányi}, \citenamefont
  {Shapeev}, \citenamefont {Thompson}, \citenamefont {Wood},\ and\
  \citenamefont {Ong}}]{SNAP2020}%
  \BibitemOpen
  \bibfield  {author} {\bibinfo {author} {\bibfnamefont {Y.}~\bibnamefont
  {Zuo}}, \bibinfo {author} {\bibfnamefont {C.}~\bibnamefont {Chen}}, \bibinfo
  {author} {\bibfnamefont {X.}~\bibnamefont {Li}}, \bibinfo {author}
  {\bibfnamefont {Z.}~\bibnamefont {Deng}}, \bibinfo {author} {\bibfnamefont
  {Y.}~\bibnamefont {Chen}}, \bibinfo {author} {\bibfnamefont {J.}~\bibnamefont
  {Behler}}, \bibinfo {author} {\bibfnamefont {G.}~\bibnamefont {Csányi}},
  \bibinfo {author} {\bibfnamefont {A.~V.}\ \bibnamefont {Shapeev}}, \bibinfo
  {author} {\bibfnamefont {A.~P.}\ \bibnamefont {Thompson}}, \bibinfo {author}
  {\bibfnamefont {M.~A.}\ \bibnamefont {Wood}}, \ and\ \bibinfo {author}
  {\bibfnamefont {S.~P.}\ \bibnamefont {Ong}},\ }\href {\doibase
  10.1021/acs.jpca.9b08723} {\bibfield  {journal} {\bibinfo  {journal} {J.
  Phys. Chem. A}\ }\textbf {\bibinfo {volume} {124}},\ \bibinfo {pages} {731}
  (\bibinfo {year} {2020})}\BibitemShut {NoStop}%
\bibitem [{\citenamefont {Unke}\ \emph {et~al.}(2021)\citenamefont {Unke},
  \citenamefont {Chmiela}, \citenamefont {Sauceda}, \citenamefont {Gastegger},
  \citenamefont {Poltavsky}, \citenamefont {Schütt}, \citenamefont
  {Tkatchenko},\ and\ \citenamefont {Müller}}]{Unke2021ChemRev}%
  \BibitemOpen
  \bibfield  {author} {\bibinfo {author} {\bibfnamefont {O.~T.}\ \bibnamefont
  {Unke}}, \bibinfo {author} {\bibfnamefont {S.}~\bibnamefont {Chmiela}},
  \bibinfo {author} {\bibfnamefont {H.~E.}\ \bibnamefont {Sauceda}}, \bibinfo
  {author} {\bibfnamefont {M.}~\bibnamefont {Gastegger}}, \bibinfo {author}
  {\bibfnamefont {I.}~\bibnamefont {Poltavsky}}, \bibinfo {author}
  {\bibfnamefont {K.~T.}\ \bibnamefont {Schütt}}, \bibinfo {author}
  {\bibfnamefont {A.}~\bibnamefont {Tkatchenko}}, \ and\ \bibinfo {author}
  {\bibfnamefont {K.-R.}\ \bibnamefont {Müller}},\ }\href {\doibase
  10.1021/acs.chemrev.0c01111} {\bibfield  {journal} {\bibinfo  {journal}
  {Chem. Rev.}\ }\textbf {\bibinfo {volume} {121}},\ \bibinfo {pages} {10142}
  (\bibinfo {year} {2021})}\BibitemShut {NoStop}%
\bibitem [{\citenamefont {Bart\'ok}, \citenamefont {Kondor},\ and\
  \citenamefont {Cs\'anyi}(2013)}]{SOAPBartok}%
  \BibitemOpen
  \bibfield  {author} {\bibinfo {author} {\bibfnamefont {A.~P.}\ \bibnamefont
  {Bart\'ok}}, \bibinfo {author} {\bibfnamefont {R.}~\bibnamefont {Kondor}}, \
  and\ \bibinfo {author} {\bibfnamefont {G.}~\bibnamefont {Cs\'anyi}},\ }\href
  {\doibase 10.1103/PhysRevB.87.184115} {\bibfield  {journal} {\bibinfo
  {journal} {Phys. Rev. B}\ }\textbf {\bibinfo {volume} {87}},\ \bibinfo
  {pages} {184115} (\bibinfo {year} {2013})}\BibitemShut {NoStop}%
\bibitem [{\citenamefont {De}\ \emph {et~al.}(2016)\citenamefont {De},
  \citenamefont {Bart{\'{o}}k}, \citenamefont {Cs{\'{a}}nyi},\ and\
  \citenamefont {Ceriotti}}]{SOAPDistance}%
  \BibitemOpen
  \bibfield  {author} {\bibinfo {author} {\bibfnamefont {S.}~\bibnamefont
  {De}}, \bibinfo {author} {\bibfnamefont {A.~P.}\ \bibnamefont
  {Bart{\'{o}}k}}, \bibinfo {author} {\bibfnamefont {G.}~\bibnamefont
  {Cs{\'{a}}nyi}}, \ and\ \bibinfo {author} {\bibfnamefont {M.}~\bibnamefont
  {Ceriotti}},\ }\href {\doibase 10.1039/C6CP00415F} {\bibfield  {journal}
  {\bibinfo  {journal} {Phys. Chem. Chem. Phys.}\ }\textbf {\bibinfo {volume}
  {18}},\ \bibinfo {pages} {13754} (\bibinfo {year} {2016})}\BibitemShut
  {NoStop}%
\bibitem [{\citenamefont {Capelli}, \citenamefont {Muniz-Miranda},\ and\
  \citenamefont {Pavan}(2022)}]{capelli2022ephemeral}%
  \BibitemOpen
  \bibfield  {author} {\bibinfo {author} {\bibfnamefont {R.}~\bibnamefont
  {Capelli}}, \bibinfo {author} {\bibfnamefont {F.}~\bibnamefont
  {Muniz-Miranda}}, \ and\ \bibinfo {author} {\bibfnamefont {G.~M.}\
  \bibnamefont {Pavan}},\ }\href@noop {} {\bibfield  {journal} {\bibinfo
  {journal} {J. Chem. Phys.}\ }\textbf {\bibinfo {volume} {156}},\ \bibinfo
  {pages} {214503} (\bibinfo {year} {2022})}\BibitemShut {NoStop}%
\bibitem [{\citenamefont {Bian}\ \emph {et~al.}(2021)\citenamefont {Bian},
  \citenamefont {Gardin}, \citenamefont {Gemen}, \citenamefont {Houben},
  \citenamefont {Perego}, \citenamefont {Lee}, \citenamefont {Elad},
  \citenamefont {Chu}, \citenamefont {Pavan},\ and\ \citenamefont
  {Klajn}}]{bian2021electrostatic}%
  \BibitemOpen
  \bibfield  {author} {\bibinfo {author} {\bibfnamefont {T.}~\bibnamefont
  {Bian}}, \bibinfo {author} {\bibfnamefont {A.}~\bibnamefont {Gardin}},
  \bibinfo {author} {\bibfnamefont {J.}~\bibnamefont {Gemen}}, \bibinfo
  {author} {\bibfnamefont {L.}~\bibnamefont {Houben}}, \bibinfo {author}
  {\bibfnamefont {C.}~\bibnamefont {Perego}}, \bibinfo {author} {\bibfnamefont
  {B.}~\bibnamefont {Lee}}, \bibinfo {author} {\bibfnamefont {N.}~\bibnamefont
  {Elad}}, \bibinfo {author} {\bibfnamefont {Z.}~\bibnamefont {Chu}}, \bibinfo
  {author} {\bibfnamefont {G.~M.}\ \bibnamefont {Pavan}}, \ and\ \bibinfo
  {author} {\bibfnamefont {R.}~\bibnamefont {Klajn}},\ }\href@noop {}
  {\bibfield  {journal} {\bibinfo  {journal} {Nat. Chem.}\ }\textbf {\bibinfo
  {volume} {13}},\ \bibinfo {pages} {940} (\bibinfo {year} {2021})}\BibitemShut
  {NoStop}%
\bibitem [{\citenamefont {Capelli}\ \emph {et~al.}(2021)\citenamefont
  {Capelli}, \citenamefont {Gardin}, \citenamefont {Empereur-Mot},
  \citenamefont {Doni},\ and\ \citenamefont {Pavan}}]{capelli2021data}%
  \BibitemOpen
  \bibfield  {author} {\bibinfo {author} {\bibfnamefont {R.}~\bibnamefont
  {Capelli}}, \bibinfo {author} {\bibfnamefont {A.}~\bibnamefont {Gardin}},
  \bibinfo {author} {\bibfnamefont {C.}~\bibnamefont {Empereur-Mot}}, \bibinfo
  {author} {\bibfnamefont {G.}~\bibnamefont {Doni}}, \ and\ \bibinfo {author}
  {\bibfnamefont {G.~M.}\ \bibnamefont {Pavan}},\ }\href@noop {} {\bibfield
  {journal} {\bibinfo  {journal} {J. Phys. Chem. B}\ }\textbf {\bibinfo
  {volume} {125}},\ \bibinfo {pages} {7785} (\bibinfo {year}
  {2021})}\BibitemShut {NoStop}%
\bibitem [{\citenamefont {Zeni}\ \emph {et~al.}(2021)\citenamefont {Zeni},
  \citenamefont {Rossi}, \citenamefont {Pavloudis}, \citenamefont {Kioseoglou},
  \citenamefont {de~Gironcoli}, \citenamefont {Palmer},\ and\ \citenamefont
  {Baletto}}]{zeni2021data}%
  \BibitemOpen
  \bibfield  {author} {\bibinfo {author} {\bibfnamefont {C.}~\bibnamefont
  {Zeni}}, \bibinfo {author} {\bibfnamefont {K.}~\bibnamefont {Rossi}},
  \bibinfo {author} {\bibfnamefont {T.}~\bibnamefont {Pavloudis}}, \bibinfo
  {author} {\bibfnamefont {J.}~\bibnamefont {Kioseoglou}}, \bibinfo {author}
  {\bibfnamefont {S.}~\bibnamefont {de~Gironcoli}}, \bibinfo {author}
  {\bibfnamefont {R.~E.}\ \bibnamefont {Palmer}}, \ and\ \bibinfo {author}
  {\bibfnamefont {F.}~\bibnamefont {Baletto}},\ }\href@noop {} {\bibfield
  {journal} {\bibinfo  {journal} {Nat. Commun.}\ }\textbf {\bibinfo {volume}
  {12}},\ \bibinfo {pages} {1} (\bibinfo {year} {2021})}\BibitemShut {NoStop}%
\bibitem [{\citenamefont {Kohn}\ and\ \citenamefont
  {Sham}(1965)}]{kohn1965self}%
  \BibitemOpen
  \bibfield  {author} {\bibinfo {author} {\bibfnamefont {W.}~\bibnamefont
  {Kohn}}\ and\ \bibinfo {author} {\bibfnamefont {L.~J.}\ \bibnamefont
  {Sham}},\ }\href@noop {} {\bibfield  {journal} {\bibinfo  {journal} {Phys.
  Rev.}\ }\textbf {\bibinfo {volume} {140}},\ \bibinfo {pages} {A1133}
  (\bibinfo {year} {1965})}\BibitemShut {NoStop}%
\bibitem [{\citenamefont {Nitopi}\ \emph {et~al.}(2019)\citenamefont {Nitopi},
  \citenamefont {Bertheussen}, \citenamefont {Scott}, \citenamefont {Liu},
  \citenamefont {Engstfeld}, \citenamefont {Horch}, \citenamefont {Seger},
  \citenamefont {Stephens}, \citenamefont {Chan}, \citenamefont {Hahn} \emph
  {et~al.}}]{nitopi2019progress}%
  \BibitemOpen
  \bibfield  {author} {\bibinfo {author} {\bibfnamefont {S.}~\bibnamefont
  {Nitopi}}, \bibinfo {author} {\bibfnamefont {E.}~\bibnamefont {Bertheussen}},
  \bibinfo {author} {\bibfnamefont {S.~B.}\ \bibnamefont {Scott}}, \bibinfo
  {author} {\bibfnamefont {X.}~\bibnamefont {Liu}}, \bibinfo {author}
  {\bibfnamefont {A.~K.}\ \bibnamefont {Engstfeld}}, \bibinfo {author}
  {\bibfnamefont {S.}~\bibnamefont {Horch}}, \bibinfo {author} {\bibfnamefont
  {B.}~\bibnamefont {Seger}}, \bibinfo {author} {\bibfnamefont {I.~E.}\
  \bibnamefont {Stephens}}, \bibinfo {author} {\bibfnamefont {K.}~\bibnamefont
  {Chan}}, \bibinfo {author} {\bibfnamefont {C.}~\bibnamefont {Hahn}},  \emph
  {et~al.},\ }\href@noop {} {\bibfield  {journal} {\bibinfo  {journal} {Chem.
  Rev.}\ }\textbf {\bibinfo {volume} {119}},\ \bibinfo {pages} {7610} (\bibinfo
  {year} {2019})}\BibitemShut {NoStop}%
\bibitem [{\citenamefont {Manthiram}, \citenamefont {Beberwyck},\ and\
  \citenamefont {Alivisatos}(2014)}]{manthiram2014enhanced}%
  \BibitemOpen
  \bibfield  {author} {\bibinfo {author} {\bibfnamefont {K.}~\bibnamefont
  {Manthiram}}, \bibinfo {author} {\bibfnamefont {B.~J.}\ \bibnamefont
  {Beberwyck}}, \ and\ \bibinfo {author} {\bibfnamefont {A.~P.}\ \bibnamefont
  {Alivisatos}},\ }\href@noop {} {\bibfield  {journal} {\bibinfo  {journal} {J.
  Am. Chem. Soc.}\ }\textbf {\bibinfo {volume} {136}},\ \bibinfo {pages}
  {13319} (\bibinfo {year} {2014})}\BibitemShut {NoStop}%
\bibitem [{\citenamefont {Hickman}\ and\ \citenamefont
  {Sanford}(2012)}]{hickman2012high}%
  \BibitemOpen
  \bibfield  {author} {\bibinfo {author} {\bibfnamefont {A.~J.}\ \bibnamefont
  {Hickman}}\ and\ \bibinfo {author} {\bibfnamefont {M.~S.}\ \bibnamefont
  {Sanford}},\ }\href@noop {} {\bibfield  {journal} {\bibinfo  {journal}
  {Nature}\ }\textbf {\bibinfo {volume} {484}},\ \bibinfo {pages} {177}
  (\bibinfo {year} {2012})}\BibitemShut {NoStop}%
\bibitem [{\citenamefont {Magdassi}, \citenamefont {Grouchko},\ and\
  \citenamefont {Kamyshny}(2010)}]{Magdassi_electronics}%
  \BibitemOpen
  \bibfield  {author} {\bibinfo {author} {\bibfnamefont {S.}~\bibnamefont
  {Magdassi}}, \bibinfo {author} {\bibfnamefont {M.}~\bibnamefont {Grouchko}},
  \ and\ \bibinfo {author} {\bibfnamefont {A.}~\bibnamefont {Kamyshny}},\
  }\href {\doibase 10.3390/ma3094626} {\bibfield  {journal} {\bibinfo
  {journal} {Materials}\ }\textbf {\bibinfo {volume} {3}},\ \bibinfo {pages}
  {4626} (\bibinfo {year} {2010})}\BibitemShut {NoStop}%
\bibitem [{\citenamefont {Lu}\ \emph {et~al.}(2004)\citenamefont {Lu},
  \citenamefont {Shen}, \citenamefont {Chen}, \citenamefont {Qian},\ and\
  \citenamefont {Lu}}]{Lu2004Electric}%
  \BibitemOpen
  \bibfield  {author} {\bibinfo {author} {\bibfnamefont {L.}~\bibnamefont
  {Lu}}, \bibinfo {author} {\bibfnamefont {Y.}~\bibnamefont {Shen}}, \bibinfo
  {author} {\bibfnamefont {X.}~\bibnamefont {Chen}}, \bibinfo {author}
  {\bibfnamefont {L.}~\bibnamefont {Qian}}, \ and\ \bibinfo {author}
  {\bibfnamefont {K.}~\bibnamefont {Lu}},\ }\href {\doibase
  10.1126/science.1092905} {\bibfield  {journal} {\bibinfo  {journal}
  {Science}\ }\textbf {\bibinfo {volume} {304}},\ \bibinfo {pages} {422}
  (\bibinfo {year} {2004})}\BibitemShut {NoStop}%
\bibitem [{\citenamefont {Behrens}\ \emph
  {et~al.}(2012{\natexlab{a}})\citenamefont {Behrens}, \citenamefont {Studt},
  \citenamefont {Kasatkin}, \citenamefont {K{\"u}hl}, \citenamefont
  {H{\"a}vecker}, \citenamefont {Abild-Pedersen}, \citenamefont {Zander},
  \citenamefont {Girgsdies}, \citenamefont {Kurr}, \citenamefont {Kniep} \emph
  {et~al.}}]{behrens2012active}%
  \BibitemOpen
  \bibfield  {author} {\bibinfo {author} {\bibfnamefont {M.}~\bibnamefont
  {Behrens}}, \bibinfo {author} {\bibfnamefont {F.}~\bibnamefont {Studt}},
  \bibinfo {author} {\bibfnamefont {I.}~\bibnamefont {Kasatkin}}, \bibinfo
  {author} {\bibfnamefont {S.}~\bibnamefont {K{\"u}hl}}, \bibinfo {author}
  {\bibfnamefont {M.}~\bibnamefont {H{\"a}vecker}}, \bibinfo {author}
  {\bibfnamefont {F.}~\bibnamefont {Abild-Pedersen}}, \bibinfo {author}
  {\bibfnamefont {S.}~\bibnamefont {Zander}}, \bibinfo {author} {\bibfnamefont
  {F.}~\bibnamefont {Girgsdies}}, \bibinfo {author} {\bibfnamefont
  {P.}~\bibnamefont {Kurr}}, \bibinfo {author} {\bibfnamefont {B.-L.}\
  \bibnamefont {Kniep}},  \emph {et~al.},\ }\href@noop {} {\bibfield  {journal}
  {\bibinfo  {journal} {Science}\ }\textbf {\bibinfo {volume} {336}},\ \bibinfo
  {pages} {893} (\bibinfo {year} {2012}{\natexlab{a}})}\BibitemShut {NoStop}%
\bibitem [{\citenamefont {Higham}, \citenamefont {Quesne},\ and\ \citenamefont
  {Catlow}(2020)}]{higham2020mechanism}%
  \BibitemOpen
  \bibfield  {author} {\bibinfo {author} {\bibfnamefont {M.~D.}\ \bibnamefont
  {Higham}}, \bibinfo {author} {\bibfnamefont {M.~G.}\ \bibnamefont {Quesne}},
  \ and\ \bibinfo {author} {\bibfnamefont {C.~R.~A.}\ \bibnamefont {Catlow}},\
  }\href@noop {} {\bibfield  {journal} {\bibinfo  {journal} {Dalton Trans.}\
  }\textbf {\bibinfo {volume} {49}},\ \bibinfo {pages} {8478} (\bibinfo {year}
  {2020})}\BibitemShut {NoStop}%
\bibitem [{\citenamefont {Gupta}(1981)}]{Gupta_energy}%
  \BibitemOpen
  \bibfield  {author} {\bibinfo {author} {\bibfnamefont {R.~P.}\ \bibnamefont
  {Gupta}},\ }\href {\doibase 10.1103/PhysRevB.23.6265} {\bibfield  {journal}
  {\bibinfo  {journal} {Phys. Rev. B}\ }\textbf {\bibinfo {volume} {23}},\
  \bibinfo {pages} {6265} (\bibinfo {year} {1981})}\BibitemShut {NoStop}%
\bibitem [{\citenamefont {Foiles}, \citenamefont {Baskes},\ and\ \citenamefont
  {Daw}(1986)}]{EAM1986}%
  \BibitemOpen
  \bibfield  {author} {\bibinfo {author} {\bibfnamefont {S.~M.}\ \bibnamefont
  {Foiles}}, \bibinfo {author} {\bibfnamefont {M.~I.}\ \bibnamefont {Baskes}},
  \ and\ \bibinfo {author} {\bibfnamefont {M.~S.}\ \bibnamefont {Daw}},\ }\href
  {\doibase 10.1103/PhysRevB.33.7983} {\bibfield  {journal} {\bibinfo
  {journal} {Phys. Rev. B}\ }\textbf {\bibinfo {volume} {33}},\ \bibinfo
  {pages} {7983} (\bibinfo {year} {1986})}\BibitemShut {NoStop}%
\bibitem [{\citenamefont {Rosato}, \citenamefont {Guillope},\ and\
  \citenamefont {Legrand}(1989)}]{Rosato1989}%
  \BibitemOpen
  \bibfield  {author} {\bibinfo {author} {\bibfnamefont {V.}~\bibnamefont
  {Rosato}}, \bibinfo {author} {\bibfnamefont {M.}~\bibnamefont {Guillope}}, \
  and\ \bibinfo {author} {\bibfnamefont {B.}~\bibnamefont {Legrand}},\ }\href
  {\doibase 10.1080/01418618908205062} {\bibfield  {journal} {\bibinfo
  {journal} {Philos. Mag. A}\ }\textbf {\bibinfo {volume} {59}},\ \bibinfo
  {pages} {321} (\bibinfo {year} {1989})}\BibitemShut {NoStop}%
\bibitem [{\citenamefont {Baskes}(1992)}]{MEAM1992}%
  \BibitemOpen
  \bibfield  {author} {\bibinfo {author} {\bibfnamefont {M.~I.}\ \bibnamefont
  {Baskes}},\ }\href {\doibase 10.1103/PhysRevB.46.2727} {\bibfield  {journal}
  {\bibinfo  {journal} {Phys. Rev. B}\ }\textbf {\bibinfo {volume} {46}},\
  \bibinfo {pages} {2727} (\bibinfo {year} {1992})}\BibitemShut {NoStop}%
\bibitem [{\citenamefont {Lee}, \citenamefont {Shim},\ and\ \citenamefont
  {Baskes}(2003)}]{MEAM2003}%
  \BibitemOpen
  \bibfield  {author} {\bibinfo {author} {\bibfnamefont {B.-J.}\ \bibnamefont
  {Lee}}, \bibinfo {author} {\bibfnamefont {J.-H.}\ \bibnamefont {Shim}}, \
  and\ \bibinfo {author} {\bibfnamefont {M.~I.}\ \bibnamefont {Baskes}},\
  }\href {\doibase 10.1103/PhysRevB.68.144112} {\bibfield  {journal} {\bibinfo
  {journal} {Phys. Rev. B}\ }\textbf {\bibinfo {volume} {68}},\ \bibinfo
  {pages} {144112} (\bibinfo {year} {2003})}\BibitemShut {NoStop}%
\bibitem [{\citenamefont {Mendelev}\ \emph {et~al.}(2008)\citenamefont
  {Mendelev}, \citenamefont {Kramer}, \citenamefont {Becker},\ and\
  \citenamefont {Asta}}]{MendelevEAM2008}%
  \BibitemOpen
  \bibfield  {author} {\bibinfo {author} {\bibfnamefont {M.}~\bibnamefont
  {Mendelev}}, \bibinfo {author} {\bibfnamefont {M.}~\bibnamefont {Kramer}},
  \bibinfo {author} {\bibfnamefont {C.}~\bibnamefont {Becker}}, \ and\ \bibinfo
  {author} {\bibfnamefont {M.}~\bibnamefont {Asta}},\ }\href {\doibase
  10.1080/14786430802206482} {\bibfield  {journal} {\bibinfo  {journal}
  {Philos. Mag.}\ }\textbf {\bibinfo {volume} {88}},\ \bibinfo {pages} {1723}
  (\bibinfo {year} {2008})}\BibitemShut {NoStop}%
\bibitem [{\citenamefont {Zhang}\ \emph {et~al.}(2018)\citenamefont {Zhang},
  \citenamefont {Han}, \citenamefont {Wang}, \citenamefont {Car},\ and\
  \citenamefont {E}}]{deepMDPRL}%
  \BibitemOpen
  \bibfield  {author} {\bibinfo {author} {\bibfnamefont {L.}~\bibnamefont
  {Zhang}}, \bibinfo {author} {\bibfnamefont {J.}~\bibnamefont {Han}}, \bibinfo
  {author} {\bibfnamefont {H.}~\bibnamefont {Wang}}, \bibinfo {author}
  {\bibfnamefont {R.}~\bibnamefont {Car}}, \ and\ \bibinfo {author}
  {\bibfnamefont {W.}~\bibnamefont {E}},\ }\href {\doibase
  10.1103/PhysRevLett.120.143001} {\bibfield  {journal} {\bibinfo  {journal}
  {Phys. Rev. Lett.}\ }\textbf {\bibinfo {volume} {120}},\ \bibinfo {pages}
  {143001} (\bibinfo {year} {2018})}\BibitemShut {NoStop}%
\bibitem [{\citenamefont {Deringer}\ and\ \citenamefont
  {Cs\'anyi}(2017)}]{Deringer2017}%
  \BibitemOpen
  \bibfield  {author} {\bibinfo {author} {\bibfnamefont {V.~L.}\ \bibnamefont
  {Deringer}}\ and\ \bibinfo {author} {\bibfnamefont {G.}~\bibnamefont
  {Cs\'anyi}},\ }\href {\doibase 10.1103/PhysRevB.95.094203} {\bibfield
  {journal} {\bibinfo  {journal} {Phys. Rev. B}\ }\textbf {\bibinfo {volume}
  {95}},\ \bibinfo {pages} {094203} (\bibinfo {year} {2017})}\BibitemShut
  {NoStop}%
\bibitem [{\citenamefont {Zhang}\ \emph {et~al.}(2020)\citenamefont {Zhang},
  \citenamefont {Wang}, \citenamefont {Chen}, \citenamefont {Zeng},
  \citenamefont {Zhang}, \citenamefont {Wang},\ and\ \citenamefont
  {E}}]{dpgen}%
  \BibitemOpen
  \bibfield  {author} {\bibinfo {author} {\bibfnamefont {Y.}~\bibnamefont
  {Zhang}}, \bibinfo {author} {\bibfnamefont {H.}~\bibnamefont {Wang}},
  \bibinfo {author} {\bibfnamefont {W.}~\bibnamefont {Chen}}, \bibinfo {author}
  {\bibfnamefont {J.}~\bibnamefont {Zeng}}, \bibinfo {author} {\bibfnamefont
  {L.}~\bibnamefont {Zhang}}, \bibinfo {author} {\bibfnamefont
  {H.}~\bibnamefont {Wang}}, \ and\ \bibinfo {author} {\bibfnamefont
  {W.}~\bibnamefont {E}},\ }\href {\doibase
  https://doi.org/10.1016/j.cpc.2020.107206} {\bibfield  {journal} {\bibinfo
  {journal} {Comput. Phys. Commun.}\ }\textbf {\bibinfo {volume} {253}},\
  \bibinfo {pages} {107206} (\bibinfo {year} {2020})}\BibitemShut {NoStop}%
\bibitem [{\citenamefont {Kittel}\ and\ \citenamefont
  {McEuen}(2018)}]{kittel2018introduction}%
  \BibitemOpen
  \bibfield  {author} {\bibinfo {author} {\bibfnamefont {C.}~\bibnamefont
  {Kittel}}\ and\ \bibinfo {author} {\bibfnamefont {P.}~\bibnamefont
  {McEuen}},\ }\href@noop {} {\emph {\bibinfo {title} {Introduction to solid
  state physics}}}\ (\bibinfo  {publisher} {John Wiley \& Sons},\ \bibinfo
  {year} {2018})\BibitemShut {NoStop}%
\bibitem [{\citenamefont {Brandes}\ and\ \citenamefont
  {Brook}(2013)}]{brandes2013smithells}%
  \BibitemOpen
  \bibfield  {author} {\bibinfo {author} {\bibfnamefont {E.~A.}\ \bibnamefont
  {Brandes}}\ and\ \bibinfo {author} {\bibfnamefont {G.}~\bibnamefont
  {Brook}},\ }\href@noop {} {\emph {\bibinfo {title} {Smithells metals
  reference book}}}\ (\bibinfo  {publisher} {Elsevier},\ \bibinfo {year}
  {2013})\BibitemShut {NoStop}%
\bibitem [{\citenamefont {Tyson}\ and\ \citenamefont
  {Miller}(1977)}]{TYSON1977267}%
  \BibitemOpen
  \bibfield  {author} {\bibinfo {author} {\bibfnamefont {W.}~\bibnamefont
  {Tyson}}\ and\ \bibinfo {author} {\bibfnamefont {W.}~\bibnamefont {Miller}},\
  }\href {\doibase https://doi.org/10.1016/0039-6028(77)90442-3} {\bibfield
  {journal} {\bibinfo  {journal} {Surface Science}\ }\textbf {\bibinfo {volume}
  {62}},\ \bibinfo {pages} {267} (\bibinfo {year} {1977})}\BibitemShut
  {NoStop}%
\bibitem [{\citenamefont {Merikoski}\ \emph {et~al.}(1997)\citenamefont
  {Merikoski}, \citenamefont {Vattulainen}, \citenamefont {Heinonen},\ and\
  \citenamefont {Ala-Nissila}}]{merikoski1997effect}%
  \BibitemOpen
  \bibfield  {author} {\bibinfo {author} {\bibfnamefont {J.}~\bibnamefont
  {Merikoski}}, \bibinfo {author} {\bibfnamefont {I.}~\bibnamefont
  {Vattulainen}}, \bibinfo {author} {\bibfnamefont {J.}~\bibnamefont
  {Heinonen}}, \ and\ \bibinfo {author} {\bibfnamefont {T.}~\bibnamefont
  {Ala-Nissila}},\ }\href@noop {} {\bibfield  {journal} {\bibinfo  {journal}
  {Surface science}\ }\textbf {\bibinfo {volume} {387}},\ \bibinfo {pages}
  {167} (\bibinfo {year} {1997})}\BibitemShut {NoStop}%
\bibitem [{\citenamefont {Merikoski}\ and\ \citenamefont
  {Ala-Nissila}(1995)}]{merikoski1995diffusion}%
  \BibitemOpen
  \bibfield  {author} {\bibinfo {author} {\bibfnamefont {J.}~\bibnamefont
  {Merikoski}}\ and\ \bibinfo {author} {\bibfnamefont {T.}~\bibnamefont
  {Ala-Nissila}},\ }\href@noop {} {\bibfield  {journal} {\bibinfo  {journal}
  {Physical Review B}\ }\textbf {\bibinfo {volume} {52}},\ \bibinfo {pages}
  {R8715} (\bibinfo {year} {1995})}\BibitemShut {NoStop}%
\bibitem [{\citenamefont {Perkins}\ and\ \citenamefont
  {DePristo}(1993)}]{perkins1993self}%
  \BibitemOpen
  \bibfield  {author} {\bibinfo {author} {\bibfnamefont {L.~S.}\ \bibnamefont
  {Perkins}}\ and\ \bibinfo {author} {\bibfnamefont {A.~E.}\ \bibnamefont
  {DePristo}},\ }\href@noop {} {\bibfield  {journal} {\bibinfo  {journal}
  {Surface science}\ }\textbf {\bibinfo {volume} {294}},\ \bibinfo {pages} {67}
  (\bibinfo {year} {1993})}\BibitemShut {NoStop}%
\bibitem [{\citenamefont {Boisvert}\ and\ \citenamefont
  {Lewis}(1997)}]{boisvert1997self}%
  \BibitemOpen
  \bibfield  {author} {\bibinfo {author} {\bibfnamefont {G.}~\bibnamefont
  {Boisvert}}\ and\ \bibinfo {author} {\bibfnamefont {L.~J.}\ \bibnamefont
  {Lewis}},\ }\href@noop {} {\bibfield  {journal} {\bibinfo  {journal}
  {Physical Review B}\ }\textbf {\bibinfo {volume} {56}},\ \bibinfo {pages}
  {7643} (\bibinfo {year} {1997})}\BibitemShut {NoStop}%
\bibitem [{\citenamefont {Liu}\ \emph {et~al.}(1991)\citenamefont {Liu},
  \citenamefont {Cohen}, \citenamefont {Adams},\ and\ \citenamefont
  {Voter}}]{liu1991eam}%
  \BibitemOpen
  \bibfield  {author} {\bibinfo {author} {\bibfnamefont {C.}~\bibnamefont
  {Liu}}, \bibinfo {author} {\bibfnamefont {J.}~\bibnamefont {Cohen}}, \bibinfo
  {author} {\bibfnamefont {J.}~\bibnamefont {Adams}}, \ and\ \bibinfo {author}
  {\bibfnamefont {A.}~\bibnamefont {Voter}},\ }\href@noop {} {\bibfield
  {journal} {\bibinfo  {journal} {Surface science}\ }\textbf {\bibinfo {volume}
  {253}},\ \bibinfo {pages} {334} (\bibinfo {year} {1991})}\BibitemShut
  {NoStop}%
\bibitem [{\citenamefont {Scheffler}(1997)}]{scheffler1997physical}%
  \BibitemOpen
  \bibfield  {author} {\bibinfo {author} {\bibfnamefont {M.}~\bibnamefont
  {Scheffler}},\ }\href@noop {} {\bibfield  {journal} {\bibinfo  {journal}
  {Physical review-series B-}\ }\textbf {\bibinfo {volume} {56}},\ \bibinfo
  {pages} {R15} (\bibinfo {year} {1997})}\BibitemShut {NoStop}%
\bibitem [{\citenamefont {D{\"u}rr}, \citenamefont {Wendelken},\ and\
  \citenamefont {Zuo}(1995)}]{durr1995island}%
  \BibitemOpen
  \bibfield  {author} {\bibinfo {author} {\bibfnamefont {H.}~\bibnamefont
  {D{\"u}rr}}, \bibinfo {author} {\bibfnamefont {J.}~\bibnamefont {Wendelken}},
  \ and\ \bibinfo {author} {\bibfnamefont {J.-K.}\ \bibnamefont {Zuo}},\
  }\href@noop {} {\bibfield  {journal} {\bibinfo  {journal} {Surface science}\
  }\textbf {\bibinfo {volume} {328}},\ \bibinfo {pages} {L527} (\bibinfo {year}
  {1995})}\BibitemShut {NoStop}%
\bibitem [{\citenamefont {Ernst}, \citenamefont {Fabre},\ and\ \citenamefont
  {Lapujoulade}(1992)}]{ernst1992nucleation}%
  \BibitemOpen
  \bibfield  {author} {\bibinfo {author} {\bibfnamefont {H.-J.}\ \bibnamefont
  {Ernst}}, \bibinfo {author} {\bibfnamefont {F.}~\bibnamefont {Fabre}}, \ and\
  \bibinfo {author} {\bibfnamefont {J.}~\bibnamefont {Lapujoulade}},\
  }\href@noop {} {\bibfield  {journal} {\bibinfo  {journal} {Physical Review
  B}\ }\textbf {\bibinfo {volume} {46}},\ \bibinfo {pages} {1929} (\bibinfo
  {year} {1992})}\BibitemShut {NoStop}%
\bibitem [{\citenamefont {Hansen}\ \emph {et~al.}(1991)\citenamefont {Hansen},
  \citenamefont {Stoltze}, \citenamefont {Jacobsen},\ and\ \citenamefont
  {N\o{}rskov}}]{Hansen1991Selfdiffusion}%
  \BibitemOpen
  \bibfield  {author} {\bibinfo {author} {\bibfnamefont {L.}~\bibnamefont
  {Hansen}}, \bibinfo {author} {\bibfnamefont {P.}~\bibnamefont {Stoltze}},
  \bibinfo {author} {\bibfnamefont {K.~W.}\ \bibnamefont {Jacobsen}}, \ and\
  \bibinfo {author} {\bibfnamefont {J.~K.}\ \bibnamefont {N\o{}rskov}},\ }\href
  {\doibase 10.1103/PhysRevB.44.6523} {\bibfield  {journal} {\bibinfo
  {journal} {Phys. Rev. B}\ }\textbf {\bibinfo {volume} {44}},\ \bibinfo
  {pages} {6523} (\bibinfo {year} {1991})}\BibitemShut {NoStop}%
\bibitem [{\citenamefont {Montalenti}\ and\ \citenamefont
  {Ferrando}(1999)}]{montalenti1999jumps}%
  \BibitemOpen
  \bibfield  {author} {\bibinfo {author} {\bibfnamefont {F.}~\bibnamefont
  {Montalenti}}\ and\ \bibinfo {author} {\bibfnamefont {R.}~\bibnamefont
  {Ferrando}},\ }\href@noop {} {\bibfield  {journal} {\bibinfo  {journal}
  {Physical Review B}\ }\textbf {\bibinfo {volume} {59}},\ \bibinfo {pages}
  {5881} (\bibinfo {year} {1999})}\BibitemShut {NoStop}%
\bibitem [{\citenamefont {Chae}, \citenamefont {Lu},\ and\ \citenamefont
  {Gustafsson}(1996)}]{Chae1996111}%
  \BibitemOpen
  \bibfield  {author} {\bibinfo {author} {\bibfnamefont {K.~H.}\ \bibnamefont
  {Chae}}, \bibinfo {author} {\bibfnamefont {H.~C.}\ \bibnamefont {Lu}}, \ and\
  \bibinfo {author} {\bibfnamefont {T.}~\bibnamefont {Gustafsson}},\ }\href
  {\doibase 10.1103/PhysRevB.54.14082} {\bibfield  {journal} {\bibinfo
  {journal} {Phys. Rev. B}\ }\textbf {\bibinfo {volume} {54}},\ \bibinfo
  {pages} {14082} (\bibinfo {year} {1996})}\BibitemShut {NoStop}%
\bibitem [{\citenamefont {Al-Rawi}, \citenamefont {Kara},\ and\ \citenamefont
  {Rahman}(2002)}]{Alrawi2002111}%
  \BibitemOpen
  \bibfield  {author} {\bibinfo {author} {\bibfnamefont {A.~N.}\ \bibnamefont
  {Al-Rawi}}, \bibinfo {author} {\bibfnamefont {A.}~\bibnamefont {Kara}}, \
  and\ \bibinfo {author} {\bibfnamefont {T.~S.}\ \bibnamefont {Rahman}},\
  }\href {\doibase 10.1103/PhysRevB.66.165439} {\bibfield  {journal} {\bibinfo
  {journal} {Phys. Rev. B}\ }\textbf {\bibinfo {volume} {66}},\ \bibinfo
  {pages} {165439} (\bibinfo {year} {2002})}\BibitemShut {NoStop}%
\bibitem [{\citenamefont {Yang}\ and\ \citenamefont
  {Rahman}(1991)}]{Yang1991Enhanced}%
  \BibitemOpen
  \bibfield  {author} {\bibinfo {author} {\bibfnamefont {L.}~\bibnamefont
  {Yang}}\ and\ \bibinfo {author} {\bibfnamefont {T.~S.}\ \bibnamefont
  {Rahman}},\ }\href {\doibase 10.1103/PhysRevLett.67.2327} {\bibfield
  {journal} {\bibinfo  {journal} {Phys. Rev. Lett.}\ }\textbf {\bibinfo
  {volume} {67}},\ \bibinfo {pages} {2327} (\bibinfo {year}
  {1991})}\BibitemShut {NoStop}%
\bibitem [{\citenamefont {Trayanov}, \citenamefont {Levi},\ and\ \citenamefont
  {Tosatti}(1989)}]{Trayanov1989missingrowrough}%
  \BibitemOpen
  \bibfield  {author} {\bibinfo {author} {\bibfnamefont {A.}~\bibnamefont
  {Trayanov}}, \bibinfo {author} {\bibfnamefont {A.~C.}\ \bibnamefont {Levi}},
  \ and\ \bibinfo {author} {\bibfnamefont {E.}~\bibnamefont {Tosatti}},\ }\href
  {\doibase 10.1209/0295-5075/8/7/013} {\bibfield  {journal} {\bibinfo
  {journal} {Europhysics Letters}\ }\textbf {\bibinfo {volume} {8}},\ \bibinfo
  {pages} {657} (\bibinfo {year} {1989})}\BibitemShut {NoStop}%
\bibitem [{\citenamefont {Behrens}\ \emph
  {et~al.}(2012{\natexlab{b}})\citenamefont {Behrens}, \citenamefont {Studt},
  \citenamefont {Kasatkin}, \citenamefont {Kühl}, \citenamefont {Hävecker},
  \citenamefont {Abild-Pedersen}, \citenamefont {Zander}, \citenamefont
  {Girgsdies}, \citenamefont {Kurr}, \citenamefont {Kniep}, \citenamefont
  {Tovar}, \citenamefont {Fischer}, \citenamefont {Nørskov},\ and\
  \citenamefont {Schlögl}}]{Science2012activesitemethanol}%
  \BibitemOpen
  \bibfield  {author} {\bibinfo {author} {\bibfnamefont {M.}~\bibnamefont
  {Behrens}}, \bibinfo {author} {\bibfnamefont {F.}~\bibnamefont {Studt}},
  \bibinfo {author} {\bibfnamefont {I.}~\bibnamefont {Kasatkin}}, \bibinfo
  {author} {\bibfnamefont {S.}~\bibnamefont {Kühl}}, \bibinfo {author}
  {\bibfnamefont {M.}~\bibnamefont {Hävecker}}, \bibinfo {author}
  {\bibfnamefont {F.}~\bibnamefont {Abild-Pedersen}}, \bibinfo {author}
  {\bibfnamefont {S.}~\bibnamefont {Zander}}, \bibinfo {author} {\bibfnamefont
  {F.}~\bibnamefont {Girgsdies}}, \bibinfo {author} {\bibfnamefont
  {P.}~\bibnamefont {Kurr}}, \bibinfo {author} {\bibfnamefont {B.-L.}\
  \bibnamefont {Kniep}}, \bibinfo {author} {\bibfnamefont {M.}~\bibnamefont
  {Tovar}}, \bibinfo {author} {\bibfnamefont {R.~W.}\ \bibnamefont {Fischer}},
  \bibinfo {author} {\bibfnamefont {J.~K.}\ \bibnamefont {Nørskov}}, \ and\
  \bibinfo {author} {\bibfnamefont {R.}~\bibnamefont {Schlögl}},\ }\href
  {\doibase 10.1126/science.1219831} {\bibfield  {journal} {\bibinfo  {journal}
  {Science}\ }\textbf {\bibinfo {volume} {336}},\ \bibinfo {pages} {893}
  (\bibinfo {year} {2012}{\natexlab{b}})}\BibitemShut {NoStop}%
\bibitem [{\citenamefont {Witte}\ \emph {et~al.}(1998)\citenamefont {Witte},
  \citenamefont {Braun}, \citenamefont {Nowack}, \citenamefont {Bartels},
  \citenamefont {Neu},\ and\ \citenamefont {Meyer}}]{Witte1998Oinduced211}%
  \BibitemOpen
  \bibfield  {author} {\bibinfo {author} {\bibfnamefont {G.}~\bibnamefont
  {Witte}}, \bibinfo {author} {\bibfnamefont {J.}~\bibnamefont {Braun}},
  \bibinfo {author} {\bibfnamefont {D.}~\bibnamefont {Nowack}}, \bibinfo
  {author} {\bibfnamefont {L.}~\bibnamefont {Bartels}}, \bibinfo {author}
  {\bibfnamefont {B.}~\bibnamefont {Neu}}, \ and\ \bibinfo {author}
  {\bibfnamefont {G.}~\bibnamefont {Meyer}},\ }\href {\doibase
  10.1103/PhysRevB.58.13224} {\bibfield  {journal} {\bibinfo  {journal} {Phys.
  Rev. B}\ }\textbf {\bibinfo {volume} {58}},\ \bibinfo {pages} {13224}
  (\bibinfo {year} {1998})}\BibitemShut {NoStop}%
\bibitem [{\citenamefont {Kirby}, \citenamefont {McKee},\ and\ \citenamefont
  {Renny}(1980)}]{Kirby1980faceting210}%
  \BibitemOpen
  \bibfield  {author} {\bibinfo {author} {\bibfnamefont {R.}~\bibnamefont
  {Kirby}}, \bibinfo {author} {\bibfnamefont {C.}~\bibnamefont {McKee}}, \ and\
  \bibinfo {author} {\bibfnamefont {L.}~\bibnamefont {Renny}},\ }\href
  {\doibase https://doi.org/10.1016/0039-6028(80)90679-2} {\bibfield  {journal}
  {\bibinfo  {journal} {Surface Science}\ }\textbf {\bibinfo {volume} {97}},\
  \bibinfo {pages} {457} (\bibinfo {year} {1980})}\BibitemShut {NoStop}%
\bibitem [{\citenamefont {Campello}, \citenamefont {Moulavi},\ and\
  \citenamefont {Sander}(2013)}]{HDBSCANCampello2013}%
  \BibitemOpen
  \bibfield  {author} {\bibinfo {author} {\bibfnamefont {R.~J.}\ \bibnamefont
  {Campello}}, \bibinfo {author} {\bibfnamefont {D.}~\bibnamefont {Moulavi}}, \
  and\ \bibinfo {author} {\bibfnamefont {J.}~\bibnamefont {Sander}},\ }\href
  {\doibase 10.1007/978-3-642-37456-2_14} {\bibfield  {journal} {\bibinfo
  {journal} {Lect. Notes Comput. Sci.}\ }\textbf {\bibinfo {volume} {7819
  LNAI}},\ \bibinfo {pages} {160} (\bibinfo {year} {2013})}\BibitemShut
  {NoStop}%
\bibitem [{\citenamefont {Campello}\ \emph {et~al.}(2015)\citenamefont
  {Campello}, \citenamefont {Moulavi}, \citenamefont {Zimek},\ and\
  \citenamefont {Sander}}]{HDBSCANStarCampello2015}%
  \BibitemOpen
  \bibfield  {author} {\bibinfo {author} {\bibfnamefont {R.~J. G.~B.}\
  \bibnamefont {Campello}}, \bibinfo {author} {\bibfnamefont {D.}~\bibnamefont
  {Moulavi}}, \bibinfo {author} {\bibfnamefont {A.}~\bibnamefont {Zimek}}, \
  and\ \bibinfo {author} {\bibfnamefont {J.}~\bibnamefont {Sander}},\ }\href
  {\doibase 10.1145/2733381} {\bibfield  {journal} {\bibinfo  {journal} {ACM
  Trans. Knowl. Discov. Data}\ }\textbf {\bibinfo {volume} {10}},\ \bibinfo
  {pages} {1} (\bibinfo {year} {2015})}\BibitemShut {NoStop}%
\bibitem [{\citenamefont {Zuckerman}(2010)}]{Zuckermann2010transmat}%
  \BibitemOpen
  \bibfield  {author} {\bibinfo {author} {\bibfnamefont {D.~M.}\ \bibnamefont
  {Zuckerman}},\ }\href@noop {} {\emph {\bibinfo {title} {Statistical Physics
  of Biomolecules: An Introduction}}}\ (\bibinfo  {publisher} {CRC Press},\
  \bibinfo {year} {2010})\BibitemShut {NoStop}%
\bibitem [{\citenamefont {Makarov}(2015)}]{Makarov2015transmat}%
  \BibitemOpen
  \bibfield  {author} {\bibinfo {author} {\bibfnamefont {D.~E.}\ \bibnamefont
  {Makarov}},\ }\href@noop {} {\emph {\bibinfo {title} {Single Molecule
  Science: Physical Principles and Models}}}\ (\bibinfo  {publisher} {CRC
  Press},\ \bibinfo {year} {2015})\BibitemShut {NoStop}%
\bibitem [{\citenamefont {Gupta}\ \emph {et~al.}(2011)\citenamefont {Gupta},
  \citenamefont {Fillmore}, \citenamefont {Jiang}, \citenamefont {Shapira},
  \citenamefont {Tao}, \citenamefont {Kuperwasser},\ and\ \citenamefont
  {Lander}}]{Gupta2011transmat}%
  \BibitemOpen
  \bibfield  {author} {\bibinfo {author} {\bibfnamefont {P.~B.}\ \bibnamefont
  {Gupta}}, \bibinfo {author} {\bibfnamefont {C.~M.}\ \bibnamefont {Fillmore}},
  \bibinfo {author} {\bibfnamefont {G.}~\bibnamefont {Jiang}}, \bibinfo
  {author} {\bibfnamefont {S.~D.}\ \bibnamefont {Shapira}}, \bibinfo {author}
  {\bibfnamefont {K.}~\bibnamefont {Tao}}, \bibinfo {author} {\bibfnamefont
  {C.}~\bibnamefont {Kuperwasser}}, \ and\ \bibinfo {author} {\bibfnamefont
  {E.~S.}\ \bibnamefont {Lander}},\ }\href {\doibase
  10.1016/j.cell.2011.07.026} {\bibfield  {journal} {\bibinfo  {journal}
  {Cell}\ }\textbf {\bibinfo {volume} {146}},\ \bibinfo {pages} {633} (\bibinfo
  {year} {2011})}\BibitemShut {NoStop}%
\bibitem [{\citenamefont {Jian-Min}, \citenamefont {Fei},\ and\ \citenamefont
  {Ke-Wei}(2004)}]{Jian_Min_2004}%
  \BibitemOpen
  \bibfield  {author} {\bibinfo {author} {\bibfnamefont {Z.}~\bibnamefont
  {Jian-Min}}, \bibinfo {author} {\bibfnamefont {M.}~\bibnamefont {Fei}}, \
  and\ \bibinfo {author} {\bibfnamefont {X.}~\bibnamefont {Ke-Wei}},\ }\href
  {\doibase 10.1088/1009-1963/13/7/020} {\bibfield  {journal} {\bibinfo
  {journal} {Chin. Phys.}\ }\textbf {\bibinfo {volume} {13}},\ \bibinfo {pages}
  {1082} (\bibinfo {year} {2004})}\BibitemShut {NoStop}%
\bibitem [{\citenamefont {Wang}\ and\ \citenamefont
  {Wang}(2014)}]{WANG2014surfene}%
  \BibitemOpen
  \bibfield  {author} {\bibinfo {author} {\bibfnamefont {J.}~\bibnamefont
  {Wang}}\ and\ \bibinfo {author} {\bibfnamefont {S.-Q.}\ \bibnamefont
  {Wang}},\ }\href {\doibase https://doi.org/10.1016/j.susc.2014.08.017}
  {\bibfield  {journal} {\bibinfo  {journal} {Surf. Sci.}\ }\textbf {\bibinfo
  {volume} {630}},\ \bibinfo {pages} {216} (\bibinfo {year}
  {2014})}\BibitemShut {NoStop}%
\bibitem [{\citenamefont {Titmuss}, \citenamefont {Wander},\ and\ \citenamefont
  {King}(1996)}]{Titmuss1996reconstruction}%
  \BibitemOpen
  \bibfield  {author} {\bibinfo {author} {\bibfnamefont {S.}~\bibnamefont
  {Titmuss}}, \bibinfo {author} {\bibfnamefont {A.}~\bibnamefont {Wander}}, \
  and\ \bibinfo {author} {\bibfnamefont {D.~A.}\ \bibnamefont {King}},\ }\href
  {\doibase 10.1021/cr950214c} {\bibfield  {journal} {\bibinfo  {journal}
  {Chem. Rev.}\ }\textbf {\bibinfo {volume} {96}},\ \bibinfo {pages} {1291}
  (\bibinfo {year} {1996})}\BibitemShut {NoStop}%
\bibitem [{\citenamefont {Koch}\ \emph {et~al.}(1992)\citenamefont {Koch},
  \citenamefont {Borbonus}, \citenamefont {Haase},\ and\ \citenamefont
  {Rieder}}]{Koch1992reconstruction}%
  \BibitemOpen
  \bibfield  {author} {\bibinfo {author} {\bibfnamefont {R.}~\bibnamefont
  {Koch}}, \bibinfo {author} {\bibfnamefont {M.}~\bibnamefont {Borbonus}},
  \bibinfo {author} {\bibfnamefont {O.}~\bibnamefont {Haase}}, \ and\ \bibinfo
  {author} {\bibfnamefont {K.~H.}\ \bibnamefont {Rieder}},\ }\href {\doibase
  10.1007/BF00348329} {\bibfield  {journal} {\bibinfo  {journal} {Appl. Phys.
  A}\ }\textbf {\bibinfo {volume} {55}},\ \bibinfo {pages} {417} (\bibinfo
  {year} {1992})}\BibitemShut {NoStop}%
\bibitem [{\citenamefont {Koch}, \citenamefont {Sturmat},\ and\ \citenamefont
  {Schulz}(2000)}]{KOCH2000reconstruction}%
  \BibitemOpen
  \bibfield  {author} {\bibinfo {author} {\bibfnamefont {R.}~\bibnamefont
  {Koch}}, \bibinfo {author} {\bibfnamefont {M.}~\bibnamefont {Sturmat}}, \
  and\ \bibinfo {author} {\bibfnamefont {J.}~\bibnamefont {Schulz}},\ }\href
  {\doibase https://doi.org/10.1016/S0039-6028(00)00260-0} {\bibfield
  {journal} {\bibinfo  {journal} {Surf. Sci.}\ }\textbf {\bibinfo {volume}
  {454-456}},\ \bibinfo {pages} {543} (\bibinfo {year} {2000})}\BibitemShut
  {NoStop}%
\bibitem [{\citenamefont {Grosse}\ \emph {et~al.}(2018)\citenamefont {Grosse},
  \citenamefont {Gao}, \citenamefont {Scholten}, \citenamefont {Sinev},
  \citenamefont {Mistry},\ and\ \citenamefont
  {Roldan~Cuenya}}]{grosse2018dynamic}%
  \BibitemOpen
  \bibfield  {author} {\bibinfo {author} {\bibfnamefont {P.}~\bibnamefont
  {Grosse}}, \bibinfo {author} {\bibfnamefont {D.}~\bibnamefont {Gao}},
  \bibinfo {author} {\bibfnamefont {F.}~\bibnamefont {Scholten}}, \bibinfo
  {author} {\bibfnamefont {I.}~\bibnamefont {Sinev}}, \bibinfo {author}
  {\bibfnamefont {H.}~\bibnamefont {Mistry}}, \ and\ \bibinfo {author}
  {\bibfnamefont {B.}~\bibnamefont {Roldan~Cuenya}},\ }\href@noop {} {\bibfield
   {journal} {\bibinfo  {journal} {Angew. Chem. Int. Ed.}\ }\textbf {\bibinfo
  {volume} {57}},\ \bibinfo {pages} {6192} (\bibinfo {year}
  {2018})}\BibitemShut {NoStop}%
\bibitem [{\citenamefont {Li}\ \emph {et~al.}(2020)\citenamefont {Li},
  \citenamefont {Kim}, \citenamefont {Louisia}, \citenamefont {Xie},
  \citenamefont {Kong}, \citenamefont {Yu}, \citenamefont {Lin}, \citenamefont
  {Aloni}, \citenamefont {Fakra},\ and\ \citenamefont
  {Yang}}]{li2020electrochemically}%
  \BibitemOpen
  \bibfield  {author} {\bibinfo {author} {\bibfnamefont {Y.}~\bibnamefont
  {Li}}, \bibinfo {author} {\bibfnamefont {D.}~\bibnamefont {Kim}}, \bibinfo
  {author} {\bibfnamefont {S.}~\bibnamefont {Louisia}}, \bibinfo {author}
  {\bibfnamefont {C.}~\bibnamefont {Xie}}, \bibinfo {author} {\bibfnamefont
  {Q.}~\bibnamefont {Kong}}, \bibinfo {author} {\bibfnamefont {S.}~\bibnamefont
  {Yu}}, \bibinfo {author} {\bibfnamefont {T.}~\bibnamefont {Lin}}, \bibinfo
  {author} {\bibfnamefont {S.}~\bibnamefont {Aloni}}, \bibinfo {author}
  {\bibfnamefont {S.~C.}\ \bibnamefont {Fakra}}, \ and\ \bibinfo {author}
  {\bibfnamefont {P.}~\bibnamefont {Yang}},\ }\href@noop {} {\bibfield
  {journal} {\bibinfo  {journal} {Proc. Natl. Acad. Sci.}\ }\textbf {\bibinfo
  {volume} {117}},\ \bibinfo {pages} {9194} (\bibinfo {year}
  {2020})}\BibitemShut {NoStop}%
\bibitem [{\citenamefont {Wang}\ \emph {et~al.}(2004)\citenamefont {Wang},
  \citenamefont {Jiang}, \citenamefont {Morikawa}, \citenamefont {Nakamura},
  \citenamefont {Cai}, \citenamefont {Pan},\ and\ \citenamefont
  {Zhao}}]{wang2004cluster}%
  \BibitemOpen
  \bibfield  {author} {\bibinfo {author} {\bibfnamefont {G.-C.}\ \bibnamefont
  {Wang}}, \bibinfo {author} {\bibfnamefont {L.}~\bibnamefont {Jiang}},
  \bibinfo {author} {\bibfnamefont {Y.}~\bibnamefont {Morikawa}}, \bibinfo
  {author} {\bibfnamefont {J.}~\bibnamefont {Nakamura}}, \bibinfo {author}
  {\bibfnamefont {Z.-S.}\ \bibnamefont {Cai}}, \bibinfo {author} {\bibfnamefont
  {Y.-M.}\ \bibnamefont {Pan}}, \ and\ \bibinfo {author} {\bibfnamefont
  {X.-Z.}\ \bibnamefont {Zhao}},\ }\href@noop {} {\bibfield  {journal}
  {\bibinfo  {journal} {Surf. Sci.}\ }\textbf {\bibinfo {volume} {570}},\
  \bibinfo {pages} {205} (\bibinfo {year} {2004})}\BibitemShut {NoStop}%
\bibitem [{\citenamefont {Delgado-Callico}\ \emph {et~al.}(2021)\citenamefont
  {Delgado-Callico}, \citenamefont {Rossi}, \citenamefont {Pinto-Miles},
  \citenamefont {Salzbrenner},\ and\ \citenamefont
  {Baletto}}]{Delgado-Callico2021}%
  \BibitemOpen
  \bibfield  {author} {\bibinfo {author} {\bibfnamefont {L.}~\bibnamefont
  {Delgado-Callico}}, \bibinfo {author} {\bibfnamefont {K.}~\bibnamefont
  {Rossi}}, \bibinfo {author} {\bibfnamefont {R.}~\bibnamefont {Pinto-Miles}},
  \bibinfo {author} {\bibfnamefont {P.}~\bibnamefont {Salzbrenner}}, \ and\
  \bibinfo {author} {\bibfnamefont {F.}~\bibnamefont {Baletto}},\ }\href
  {\doibase 10.1039/D0NR06850K} {\bibfield  {journal} {\bibinfo  {journal}
  {Nanoscale}\ }\textbf {\bibinfo {volume} {13}},\ \bibinfo {pages} {1172}
  (\bibinfo {year} {2021})}\BibitemShut {NoStop}%
\bibitem [{\citenamefont {Giannozzi}\ \emph {et~al.}(2009)\citenamefont
  {Giannozzi}, \citenamefont {Baroni}, \citenamefont {Bonini}, \citenamefont
  {Calandra}, \citenamefont {Car}, \citenamefont {Cavazzoni}, \citenamefont
  {Ceresoli}, \citenamefont {Chiarotti}, \citenamefont {Cococcioni},
  \citenamefont {Dabo} \emph {et~al.}}]{giannozzi2009quantum}%
  \BibitemOpen
  \bibfield  {author} {\bibinfo {author} {\bibfnamefont {P.}~\bibnamefont
  {Giannozzi}}, \bibinfo {author} {\bibfnamefont {S.}~\bibnamefont {Baroni}},
  \bibinfo {author} {\bibfnamefont {N.}~\bibnamefont {Bonini}}, \bibinfo
  {author} {\bibfnamefont {M.}~\bibnamefont {Calandra}}, \bibinfo {author}
  {\bibfnamefont {R.}~\bibnamefont {Car}}, \bibinfo {author} {\bibfnamefont
  {C.}~\bibnamefont {Cavazzoni}}, \bibinfo {author} {\bibfnamefont
  {D.}~\bibnamefont {Ceresoli}}, \bibinfo {author} {\bibfnamefont {G.~L.}\
  \bibnamefont {Chiarotti}}, \bibinfo {author} {\bibfnamefont {M.}~\bibnamefont
  {Cococcioni}}, \bibinfo {author} {\bibfnamefont {I.}~\bibnamefont {Dabo}},
  \emph {et~al.},\ }\href@noop {} {\bibfield  {journal} {\bibinfo  {journal}
  {J. Condens. Matter Phys.}\ }\textbf {\bibinfo {volume} {21}},\ \bibinfo
  {pages} {395502} (\bibinfo {year} {2009})}\BibitemShut {NoStop}%
\bibitem [{\citenamefont {Perdew}, \citenamefont {Burke},\ and\ \citenamefont
  {Ernzerhof}(1996)}]{perdew1996generalized}%
  \BibitemOpen
  \bibfield  {author} {\bibinfo {author} {\bibfnamefont {J.~P.}\ \bibnamefont
  {Perdew}}, \bibinfo {author} {\bibfnamefont {K.}~\bibnamefont {Burke}}, \
  and\ \bibinfo {author} {\bibfnamefont {M.}~\bibnamefont {Ernzerhof}},\
  }\href@noop {} {\bibfield  {journal} {\bibinfo  {journal} {Phys. Rev. Lett.}\
  }\textbf {\bibinfo {volume} {77}},\ \bibinfo {pages} {3865} (\bibinfo {year}
  {1996})}\BibitemShut {NoStop}%
\bibitem [{\citenamefont {Andolina}\ \emph {et~al.}(2021)\citenamefont
  {Andolina}, \citenamefont {Bon}, \citenamefont {Passerone},\ and\
  \citenamefont {Saidi}}]{andolina-AuAgNP}%
  \BibitemOpen
  \bibfield  {author} {\bibinfo {author} {\bibfnamefont {C.~M.}\ \bibnamefont
  {Andolina}}, \bibinfo {author} {\bibfnamefont {M.}~\bibnamefont {Bon}},
  \bibinfo {author} {\bibfnamefont {D.}~\bibnamefont {Passerone}}, \ and\
  \bibinfo {author} {\bibfnamefont {W.~A.}\ \bibnamefont {Saidi}},\ }\href
  {\doibase 10.1021/acs.jpcc.1c04403} {\bibfield  {journal} {\bibinfo
  {journal} {J. Phys. Chem. C}\ }\textbf {\bibinfo {volume} {125}},\ \bibinfo
  {pages} {17438} (\bibinfo {year} {2021})}\BibitemShut {NoStop}%
\bibitem [{\citenamefont {Rappe}\ \emph {et~al.}(1990)\citenamefont {Rappe},
  \citenamefont {Rabe}, \citenamefont {Kaxiras},\ and\ \citenamefont
  {Joannopoulos}}]{rappe1990optimized}%
  \BibitemOpen
  \bibfield  {author} {\bibinfo {author} {\bibfnamefont {A.~M.}\ \bibnamefont
  {Rappe}}, \bibinfo {author} {\bibfnamefont {K.~M.}\ \bibnamefont {Rabe}},
  \bibinfo {author} {\bibfnamefont {E.}~\bibnamefont {Kaxiras}}, \ and\
  \bibinfo {author} {\bibfnamefont {J.}~\bibnamefont {Joannopoulos}},\
  }\href@noop {} {\bibfield  {journal} {\bibinfo  {journal} {Phys. Rev. B}\
  }\textbf {\bibinfo {volume} {41}},\ \bibinfo {pages} {1227} (\bibinfo {year}
  {1990})}\BibitemShut {NoStop}%
\bibitem [{\citenamefont {Marzari}\ \emph {et~al.}(1999)\citenamefont
  {Marzari}, \citenamefont {Vanderbilt}, \citenamefont {Vita},\ and\
  \citenamefont {Payne}}]{marzari1999thermal}%
  \BibitemOpen
  \bibfield  {author} {\bibinfo {author} {\bibfnamefont {N.}~\bibnamefont
  {Marzari}}, \bibinfo {author} {\bibfnamefont {D.}~\bibnamefont {Vanderbilt}},
  \bibinfo {author} {\bibfnamefont {A.~D.}\ \bibnamefont {Vita}}, \ and\
  \bibinfo {author} {\bibfnamefont {M.~C.}\ \bibnamefont {Payne}},\ }\href
  {\doibase 10.1103/PhysRevLett.82.3296} {\bibfield  {journal} {\bibinfo
  {journal} {Phys. Rev. Lett.}\ }\textbf {\bibinfo {volume} {82}},\ \bibinfo
  {pages} {3296} (\bibinfo {year} {1999})}\BibitemShut {NoStop}%
\bibitem [{\citenamefont {Monkhorst}\ and\ \citenamefont
  {Pack}(1976)}]{monkhorst1976special}%
  \BibitemOpen
  \bibfield  {author} {\bibinfo {author} {\bibfnamefont {H.~J.}\ \bibnamefont
  {Monkhorst}}\ and\ \bibinfo {author} {\bibfnamefont {J.~D.}\ \bibnamefont
  {Pack}},\ }\href@noop {} {\bibfield  {journal} {\bibinfo  {journal} {Phys.
  Rev. B}\ }\textbf {\bibinfo {volume} {13}},\ \bibinfo {pages} {5188}
  (\bibinfo {year} {1976})}\BibitemShut {NoStop}%
\bibitem [{\citenamefont {Bussi}, \citenamefont {Donadio},\ and\ \citenamefont
  {Parrinello}(2007)}]{bussi2007canonical}%
  \BibitemOpen
  \bibfield  {author} {\bibinfo {author} {\bibfnamefont {G.}~\bibnamefont
  {Bussi}}, \bibinfo {author} {\bibfnamefont {D.}~\bibnamefont {Donadio}}, \
  and\ \bibinfo {author} {\bibfnamefont {M.}~\bibnamefont {Parrinello}},\
  }\href@noop {} {\bibfield  {journal} {\bibinfo  {journal} {J. Chem. Phys.}\
  }\textbf {\bibinfo {volume} {126}},\ \bibinfo {pages} {014101} (\bibinfo
  {year} {2007})}\BibitemShut {NoStop}%
\bibitem [{\citenamefont {Kingma}\ and\ \citenamefont
  {Ba}(2014)}]{kingma2014adam}%
  \BibitemOpen
  \bibfield  {author} {\bibinfo {author} {\bibfnamefont {D.~P.}\ \bibnamefont
  {Kingma}}\ and\ \bibinfo {author} {\bibfnamefont {J.}~\bibnamefont {Ba}},\
  }\href@noop {} {\bibfield  {journal} {\bibinfo  {journal} {arXiv preprint
  arXiv:1412.6980}\ } (\bibinfo {year} {2014})}\BibitemShut {NoStop}%
\bibitem [{\citenamefont {Cyrot-Lackmann}\ and\ \citenamefont
  {Ducastelle}(1971)}]{CyrotDucastelle_bindingTransition}%
  \BibitemOpen
  \bibfield  {author} {\bibinfo {author} {\bibfnamefont {F.}~\bibnamefont
  {Cyrot-Lackmann}}\ and\ \bibinfo {author} {\bibfnamefont {F.}~\bibnamefont
  {Ducastelle}},\ }\href {\doibase 10.1103/PhysRevB.4.2406} {\bibfield
  {journal} {\bibinfo  {journal} {Phys. Rev. B}\ }\textbf {\bibinfo {volume}
  {4}},\ \bibinfo {pages} {2406} (\bibinfo {year} {1971})}\BibitemShut
  {NoStop}%
\bibitem [{\citenamefont {Thompson}\ \emph {et~al.}(2022)\citenamefont
  {Thompson}, \citenamefont {Aktulga}, \citenamefont {Berger}, \citenamefont
  {Bolintineanu}, \citenamefont {Brown}, \citenamefont {Crozier}, \citenamefont
  {{in 't Veld}}, \citenamefont {Kohlmeyer}, \citenamefont {Moore},
  \citenamefont {Nguyen}, \citenamefont {Shan}, \citenamefont {Stevens},
  \citenamefont {Tranchida}, \citenamefont {Trott},\ and\ \citenamefont
  {Plimpton}}]{LAMMPS}%
  \BibitemOpen
  \bibfield  {author} {\bibinfo {author} {\bibfnamefont {A.~P.}\ \bibnamefont
  {Thompson}}, \bibinfo {author} {\bibfnamefont {H.~M.}\ \bibnamefont
  {Aktulga}}, \bibinfo {author} {\bibfnamefont {R.}~\bibnamefont {Berger}},
  \bibinfo {author} {\bibfnamefont {D.~S.}\ \bibnamefont {Bolintineanu}},
  \bibinfo {author} {\bibfnamefont {W.~M.}\ \bibnamefont {Brown}}, \bibinfo
  {author} {\bibfnamefont {P.~S.}\ \bibnamefont {Crozier}}, \bibinfo {author}
  {\bibfnamefont {P.~J.}\ \bibnamefont {{in 't Veld}}}, \bibinfo {author}
  {\bibfnamefont {A.}~\bibnamefont {Kohlmeyer}}, \bibinfo {author}
  {\bibfnamefont {S.~G.}\ \bibnamefont {Moore}}, \bibinfo {author}
  {\bibfnamefont {T.~D.}\ \bibnamefont {Nguyen}}, \bibinfo {author}
  {\bibfnamefont {R.}~\bibnamefont {Shan}}, \bibinfo {author} {\bibfnamefont
  {M.~J.}\ \bibnamefont {Stevens}}, \bibinfo {author} {\bibfnamefont
  {J.}~\bibnamefont {Tranchida}}, \bibinfo {author} {\bibfnamefont
  {C.}~\bibnamefont {Trott}}, \ and\ \bibinfo {author} {\bibfnamefont {S.~J.}\
  \bibnamefont {Plimpton}},\ }\href {\doibase 10.1016/j.cpc.2021.108171}
  {\bibfield  {journal} {\bibinfo  {journal} {Comput. Phys. Commun.}\ }\textbf
  {\bibinfo {volume} {271}},\ \bibinfo {pages} {108171} (\bibinfo {year}
  {2022})}\BibitemShut {NoStop}%
\bibitem [{\citenamefont {Stukowski}(2009)}]{stukowski2009visualization}%
  \BibitemOpen
  \bibfield  {author} {\bibinfo {author} {\bibfnamefont {A.}~\bibnamefont
  {Stukowski}},\ }\href@noop {} {\bibfield  {journal} {\bibinfo  {journal}
  {Model. Simul. Mater. Sci. Eng.}\ }\textbf {\bibinfo {volume} {18}},\
  \bibinfo {pages} {015012} (\bibinfo {year} {2009})}\BibitemShut {NoStop}%
\bibitem [{\citenamefont {Himanen}\ \emph {et~al.}(2020)\citenamefont
  {Himanen}, \citenamefont {J{\"a}ger}, \citenamefont {Morooka}, \citenamefont
  {Canova}, \citenamefont {Ranawat}, \citenamefont {Gao}, \citenamefont
  {Rinke},\ and\ \citenamefont {Foster}}]{himanen2020dscribe}%
  \BibitemOpen
  \bibfield  {author} {\bibinfo {author} {\bibfnamefont {L.}~\bibnamefont
  {Himanen}}, \bibinfo {author} {\bibfnamefont {M.~O.}\ \bibnamefont
  {J{\"a}ger}}, \bibinfo {author} {\bibfnamefont {E.~V.}\ \bibnamefont
  {Morooka}}, \bibinfo {author} {\bibfnamefont {F.~F.}\ \bibnamefont {Canova}},
  \bibinfo {author} {\bibfnamefont {Y.~S.}\ \bibnamefont {Ranawat}}, \bibinfo
  {author} {\bibfnamefont {D.~Z.}\ \bibnamefont {Gao}}, \bibinfo {author}
  {\bibfnamefont {P.}~\bibnamefont {Rinke}}, \ and\ \bibinfo {author}
  {\bibfnamefont {A.~S.}\ \bibnamefont {Foster}},\ }\href@noop {} {\bibfield
  {journal} {\bibinfo  {journal} {Comput. Phys. Commun.}\ }\textbf {\bibinfo
  {volume} {247}},\ \bibinfo {pages} {106949} (\bibinfo {year}
  {2020})}\BibitemShut {NoStop}%
\bibitem [{\citenamefont {Pedregosa}\ \emph {et~al.}(2011)\citenamefont
  {Pedregosa}, \citenamefont {Varoquaux}, \citenamefont {Gramfort},
  \citenamefont {Michel}, \citenamefont {Thirion}, \citenamefont {Grisel},
  \citenamefont {Blondel}, \citenamefont {Prettenhofer}, \citenamefont {Weiss},
  \citenamefont {Dubourg}, \citenamefont {Vanderplas}, \citenamefont {Passos},
  \citenamefont {Cournapeau}, \citenamefont {Brucher}, \citenamefont {Perrot},\
  and\ \citenamefont {Duchesnay}}]{scikit-learn}%
  \BibitemOpen
  \bibfield  {author} {\bibinfo {author} {\bibfnamefont {F.}~\bibnamefont
  {Pedregosa}}, \bibinfo {author} {\bibfnamefont {G.}~\bibnamefont
  {Varoquaux}}, \bibinfo {author} {\bibfnamefont {A.}~\bibnamefont {Gramfort}},
  \bibinfo {author} {\bibfnamefont {V.}~\bibnamefont {Michel}}, \bibinfo
  {author} {\bibfnamefont {B.}~\bibnamefont {Thirion}}, \bibinfo {author}
  {\bibfnamefont {O.}~\bibnamefont {Grisel}}, \bibinfo {author} {\bibfnamefont
  {M.}~\bibnamefont {Blondel}}, \bibinfo {author} {\bibfnamefont
  {P.}~\bibnamefont {Prettenhofer}}, \bibinfo {author} {\bibfnamefont
  {R.}~\bibnamefont {Weiss}}, \bibinfo {author} {\bibfnamefont
  {V.}~\bibnamefont {Dubourg}}, \bibinfo {author} {\bibfnamefont
  {J.}~\bibnamefont {Vanderplas}}, \bibinfo {author} {\bibfnamefont
  {A.}~\bibnamefont {Passos}}, \bibinfo {author} {\bibfnamefont
  {D.}~\bibnamefont {Cournapeau}}, \bibinfo {author} {\bibfnamefont
  {M.}~\bibnamefont {Brucher}}, \bibinfo {author} {\bibfnamefont
  {M.}~\bibnamefont {Perrot}}, \ and\ \bibinfo {author} {\bibfnamefont
  {E.}~\bibnamefont {Duchesnay}},\ }\href@noop {} {\bibfield  {journal}
  {\bibinfo  {journal} {J. Mach. Learn. Res.}\ }\textbf {\bibinfo {volume}
  {12}},\ \bibinfo {pages} {2825} (\bibinfo {year} {2011})}\BibitemShut
  {NoStop}%
\end{thebibliography}%

\end{document}